\PassOptionsToPackage{pdfa,bookmarks,colorlinks=true,pagebackref=true,backref=page}{hyperref}
\PassOptionsToPackage{hyphens}{url} % acmart compatibility
\documentclass[sigconf,authorversion,nonacm]{acmart}

\input{glyphtounicode.tex}
\input{glyphtounicode-cmr.tex}
\usepackage{cmap}
\hypersetup{%
    unicode,
    pdfapart=2,
    pdfaconformance=B
}

\usepackage[vlined,linesnumbered,ruled,noend]{algorithm2e} % algorithms
\usepackage[hyperpageref]{backref} % backreferences
\usepackage{balance} % to better equalize the last page
\usepackage{bold-extra} % bold + {small capital, italic}
\usepackage{booktabs} % nicer tables
\usepackage[font={bf}, tableposition=top]{caption} % captions on top for tables
\usepackage{cleveref}
\crefname{algorithm}{Alg.}{Algs.}
\Crefname{algorithm}{Algorithm}{Algorithms}
\crefname{appendix}{App.}{App.}
\Crefname{appendix}{Appendix}{Appendices}
\crefname{corollary}{Corol.}{Corolls.}
\Crefname{corollary}{Corollary}{Corollaries}
\crefname{conjecture}{Conjecture}{Conjectures}
\Crefname{conjecture}{Conjecture}{Conjectures}
\crefformat{conjecture}{Conjecture~#2#1#3}
\crefmultiformat{conjecture}{Conjectures~#2#1#3}{\ and~#2#1#3}{, #2#1#3}{\ and~#2#1#3}
\crefname{definition}{Def.}{Defs.}
\Crefname{definition}{Definition}{Definition}
\crefname{fact}{Fact}{Facts }
\Crefname{figure}{Fact}{Facts}
\crefname{figure}{Fig.}{Figs.}
\Crefname{figure}{Figure}{Figures}
\crefname{lemma}{Lemma}{Lemmas}
\Crefname{lemma}{Lemma}{Lemmas}
\crefname{proposition}{Prop.}{Props.}
\Crefname{proposition}{Proposition}{Propositions}
\Crefname{section}{Section}{Sections}
\crefname{section}{Sect.}{Sect.}
\crefname{subsection}{Sect.}{Sect.}
\Crefname{subsection}{Section}{Sections}
\crefname{subsubsection}{Sect.}{Sect.}
\Crefname{subsubsection}{Section}{Sections}
\crefname{table}{Table}{Tables}
\Crefname{table}{Table}{Tables}
\crefname{theorem}{Thm.}{Thms.}
\Crefname{theorem}{Theorem}{Theorems}
\usepackage{enumitem}
\usepackage{graphicx} % advanced figures
\graphicspath{{figures/}{img/}}
\PassOptionsToPackage{bookmarks, pdftex, colorlinks=true, pagebackref=true, backref=page}{hyperref} % acmart compatibility
\usepackage{hyphenat} % name in single line
\usepackage[font=small,labelfont=bf,labelsep=colon,tableposition=top]{caption}
\usepackage[font=small,labelfont=bf]{subcaption} % subfloats
\usepackage{mathtools} % amsmath++
\usepackage{microtype} % compress text
\usepackage{multirow} % multi rows for tables
\PassOptionsToPackage{square,numbers}{natbib} % acmart compatibility
\usepackage{siunitx} % \num for decimal grouping
\usepackage{subcaption} % subfloats
\usepackage{type1cm} % type1 computer modern font
\PassOptionsToPackage{hyphens}{url} % acmart compatibility
\usepackage{xfrac} % nicer slanted fractions
\usepackage{xspace} % fix space in macros

\usepackage[hide]{chato-notes} % notes

\AtEndPreamble{%
\theoremstyle{acmplain}
\newtheorem{fact}[theorem]{Fact}%
}

\newcommand{\spara}[1]{\vspace{1mm}\noindent\textbf{#1}}
\newcommand{\mpara}[1]{\medskip\noindent\textbf{#1}}

\newenvironment{squishlist}
{\begin{list}{$\bullet$}
 {\setlength{\itemsep}{0pt}
     \setlength{\parsep}{3pt}
     \setlength{\topsep}{3pt}
     \setlength{\partopsep}{0pt}
     \setlength{\leftmargin}{1.5em}
     \setlength{\labelwidth}{1em}
     \setlength{\labelsep}{0.5em} } }
{\end{list}}

\renewcommand*\backref[1]{\ifx#1\relax \else (Cited on #1) \fi}

\DeclarePairedDelimiter\abs{\lvert}{\rvert} % absolute value
\makeatletter
\let\oldabs\abs%
\def\abs{\@ifstar{\oldabs}{\oldabs*}}
\makeatother

\newcommand{\acceptprob}[2]{\ensuremath{\alpha_{#1}(#2)}\xspace} % acceptance probability of neighbor #2 from state #1
\newcommand{\algonamecollective}{\algoname-{\raisebox{-1pt}{\resizebox{7pt}{!}{\ensuremath{\ast}}}}\xspace}
\newcommand{\algonamecollectiveplain}{\algoname-*} % collective name in plain text
\newcommand{\algonameplain}{Polaris\xspace} % algorithm name in plain text
\newcommand{\algoname}{\textsc{\algonameplain}\xspace} % algorithm name in smallcaps
\newcommand{\algofosdick}{\textsc{\algonameplain-B}\xspace} % Fosdick-based algorithm name
\newcommand{\algomulti}{\textsc{\algonameplain-C}\xspace} % Multiset-sampling algorithm name
\newcommand{\card}[1]{\ensuremath{\abs{#1}}\xspace} % cardinality of a set
\newcommand{\degree}[2]{\ensuremath{d_{#2}(#1)}\xspace} % degree of #1 in #2
\newcommand{\jcm}[1]{\ensuremath{J_{#1}}\xspace} % JLM of #1
\newcommand{\labelset}{\ensuremath{\mathcal{L}}\xspace} % set of labels
\newcommand{\labelfsym}{\ensuremath{\lambda}} % symbol for the labeling function
\newcommand{\labelf}[1]{\ensuremath{\labelfsym(#1)}\xspace} % label of vertex #1
\newcommand{\ldblbrace}{\{\mskip-5mu\{} % double left curvy bracket
\newcommand{\rdblbrace}{\}\mskip-5mu\}} % double right curvy bracket
\DeclarePairedDelimiter\mset{\ldblbrace}{\rdblbrace} % multiset
\newcommand{\mul}[2]{\ensuremath{\omega_{#2}(#1)}\xspace} % multiplicity of #1 in multiset #2
\newcommand{\lneighs}[3]{\ensuremath{\Gamma^{#3}_{#2}(#1)}\xspace} % neighbors of #1 in #2 with label #3
\newcommand{\nlneighs}[3]{\ensuremath{\gamma^{#3}_{#2}(#1)}\xspace} % number of neighbors of #1 in #2 with label #3
\newcommand{\neighs}[2]{\ensuremath{\Gamma_{#2}(#1)}\xspace} % neighbors of #1 in #2
\newcommand{\neighprobsym}[1]{\ensuremath{\xi_{#1}}} % symbol for the neighbor proposal probability from state #1
\newcommand{\neighprob}[2]{\ensuremath{\neighprobsym{#1}(#2)}\xspace} % neighbor proposal probability from state #1 to neighbor #2
\newcommand{\nullmodel}{\ensuremath{\Pi}\xspace} % null model
\newcommand{\nullprob}{\ensuremath{\pi}\xspace} % null probability distribution
\newcommand{\nullset}{\ensuremath{\mathcal{Z}}\xspace} % null set
\newcommand{\observed}{\ensuremath{\mathring{G}}\xspace} % observed graph
\newcommand{\props}{\ensuremath{\mathcal{P}}\xspace} % properties to be preserved
\DeclarePairedDelimiter\set{\{}{\}} % set
\newcommand{\setf}[1]{\ensuremath{\mathsf{set}(#1)}\xspace} % set version of multiset #1
\newcommand{\suchthat}{\ensuremath{\mathrel{:}}} % "such that" symbol for set definitions, with the right spacing
\newcommand{\statprob}{\ensuremath{\phi}\xspace} % stationary distribution
\newcommand{\states}{\ensuremath{\mathcal{S}}\xspace} % states in a Markov chain
\newcommand{\swap}[2]{\ensuremath{#1 \rightarrow #2}} % DES involving edges #1 to edges #2
\newcommand{\titlefirstpart}{Sampling from the Multigraph Configuration Model}
\newcommand{\titlesecondpart}{with Prescribed Color Assortativity}
\newcommand{\shortertitle}{\titlefirstpart\ \titlesecondpart}
\newcommand{\transprob}[2]{\ensuremath{\tau_{#1,#2}}} % transition probability from state #1 to state #2
\newcommand{\samplespace}{\ensuremath{\mathcal{G}}\xspace} % sample space of MCMC

\newcommand{\rev}[1]{{\color{black} #1}}

\begin{document}

%\title{Polaris: A Null Model for Colored Multi-Graphs}
%\title{Polaris: A Null Model for Statistical Analysis of Colored Multi-Graphs}
%\title{Polaris: A Multi-Graph Configuration Model with Color Assortativity Constraint}
%\title{Polaris: A Null Model for Polarized Networks}
%\title{Polaris: A Network Null Model that Preserves Label Assortativity}
%\title{Polaris: An Ensemble for Multi-Graph with Fixed Color Assortativity}
%\title{Polaris: A Null Model for Multigraphs with Color Assortativity Constraints}
%\title{Polaris: Efficient Sampling of Random Multigraphs with Prescribed Color Assortativity}
%\title{Polaris: Efficient Sampling from Multigraph Ensembles with Prescribed Color Assortativity}
\title[\shortertitle]{\texorpdfstring{\algoname: \titlefirstpart\\\titlesecondpart}{\algoname: \titlefirstpart\
\titlesecondpart}}

 \author{Giulia Preti}
 \orcid{https://orcid.org/0000-0002-2126-326X}
 \email{giulia.preti@centai.eu}
 \affiliation{%
   \institution{CENTAI}
   \city{Turin}
   \country{Italy}}
   
 \author{Matteo Riondato}
 \orcid{0000-0003-2523-4420}
 \email{mriondato@amherst.edu}
 \affiliation{%
   \institution{Amherst College}
   \city{Amherst, MA}
   \country{USA}}
   
 \author{Aristides Gionis}
 \orcid{https://orcid.org/0000-0002-5211-112X}
 \email{argioni@kth.se}
 \affiliation{%
   \institution{KTH Royal Institute of Technology}
   \city{Stockholm}
   \country{Sweden}}

 \author{Gianmarco De Francisci Morales}
 \orcid{https://orcid.org/0000-0002-2415-494X}
 \email{gdfm@acm.org}
 \affiliation{%
   \institution{CENTAI}
   \city{Turin}
   \country{Italy}}

\begin{abstract}
We introduce \algoname, a network null model for colored multigraphs that preserves the Joint Color Matrix.
\algoname is specifically designed for studying network polarization, where vertices belong to a side in a debate or a partisan group, represented by a vertex color, and relations have different strengths, represented by an integer-valued edge multiplicity.
The key feature of \algoname is preserving the Joint Color Matrix (JCM) of the multigraph, which specifies the number of edges connecting vertices of any two given colors.
The JCM is the basic property that determines color assortativity, a fundamental aspect in studying homophily and segregation in polarized networks.
By using \algoname, network scientists can test whether a phenomenon is entirely explained by the JCM of the observed network or whether other phenomena might be at play.

Technically, our null model is an extension of the configuration model: an ensemble of colored multigraphs characterized by the same degree sequence and the same JCM.%
To sample from this ensemble, we develop a suite of Markov Chain Monte Carlo algorithms, collectively named \algonamecollective.
It includes \algofosdick, an adaptation of a generic Metropolis-Hastings algorithm, and \algomulti, a faster, specialized algorithm with higher acceptance probabilities.
This new null model and the associated algorithms provide a more nuanced toolset for examining polarization in social networks, thus enabling statistically sound conclusions.
\end{abstract}



% Separate the keywords with commas.

\maketitle

% !TEX root =  ../main.tex
\section{Introduction}\label{sec:intro}

\begin{figure}[t!]
  \centering
  \includegraphics[width=\linewidth]{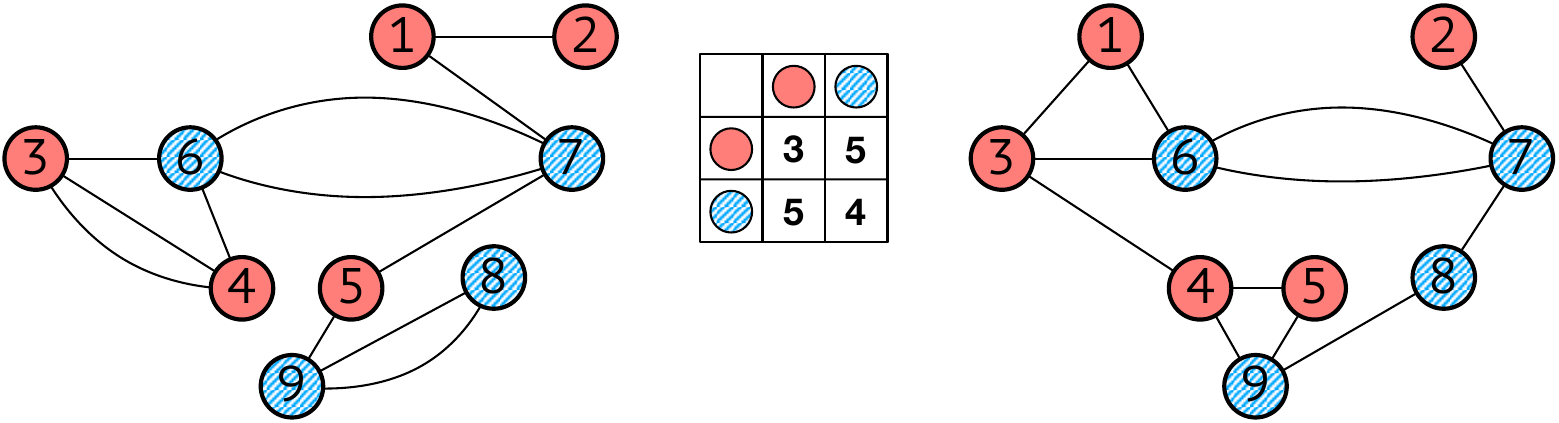}
  \caption{Two multigraphs with the same degree sequence and JCM.}%
  \Description{Examples of two multigraphs with the same degree sequence and
    JCM, and the actual JCM.}%
  \label{fig:lab_graphs}
\end{figure}

% Polarization in social networks
Polarization is perceived as one of the largest problems in our society~\citep{worldeconomicforum2024global}.
Scientists have studied the phenomenon extensively, more recently by using data from social media~\cite{conover2011political,hohmann2023quantifying,nyhan2023likeminded,guess2023reshares,gonzalez-bailon2023asymmetric}.
Many different theories try to explain the phenomenon, from affective polarization to partisan identity and echo chambers~\cite{gentzkow2011ideological,messing2014selective,garimella2018political,cinelli2021echo,baldassarri2021emergence,tornberg2022how}.
However, definite evidence is still lacking.

% Need for statistically sound results: null models
Existing observational studies often use network representations to study the problem.
This choice allows employing the ample network- and graph-theoretical toolset to define properties and compute relevant measures.
However, such quantities are only significant insofar as they are not statistical noise.
For this purpose, \emph{null-hypothesis models} are used to assess statistical significance.
% is one of the main tools.

% Lack of null models that deal with polarization
Unfortunately, to date, the network null models used in these studies are exceedingly simple~\citep{salloum2022separating}.
They usually preserve only basic characteristics of the graph structure, such as density or degree sequences~\citep{fosdick2018configuring}, 
but they ignore the interplay bet\-ween opinions or communities that describe the polarization~phenomenon.
%  and the graph structure on which communication happens.

% Our contribution: null model that preserves color assortativity in multigraphs (more general)
The main contribution of this work is to propose a new network null model geared towards the study of network polarization.
In particular, our null model is a \emph{statistical ensemble of colored
multigraphs}: graphs where each vertex has a color (i.e., a single label) and edges can appear multiple times.
% Multigraphs are a natural representation of endorsement networks (e.g., retweet network)
Vertex colors, or labels, are often used to represent the different sides in a controversial argument or debate, or groups such as partisan identities~\citep{conover2011political,garimella2016quantifying,garimella2018quantifying}.
Multiedges are commonly used to represent endorsement networks (e.g., retweet or interaction networks)~\citep{garimella2016quantifying,garimella2018quantifying,garimella2017reducing}, where the multiplicity represents the strength of the relationship between two vertices.

% Preserving JCM
The ensemble we consider is a microcanonical one akin to the configuration model~\citep{bender1978asymptotic,bollobas1980probabilistic}, i.e., its members are all and only the graphs with a specific degree sequence.
The graph ensemble for our null model is additionally defined by a property shared by all members of the ensemble: the Joint Color Matrix (JCM).
This matrix determines the number of edges that connect vertices of different colors.
\Cref{fig:lab_graphs} depicts two small graphs belonging to the same ensemble and their associated JCM.
The JCM determines important properties of the graph, e.g., its color assortativity~\citep{newman2003mixing} which is fundamental in the study of homophily and segregation~\citep{mcpherson2001birds}.

% MCMC algorithms to sample from microcanonical ensemble
We devise a suite of Markov chain Monte Carlo algorithms, named \algonamecollective, to sample from the ensemble.
We prove the Markov chain is irreducible and aperiodic, thus having a unique stationary distribution.
The first algorithm, \algofosdick, is an adaptation of an existing
algorithm~\citep{fosdick2017configuring} using the Metropolis-Hastings method.
The second algorithm, \algomulti, % is tailored to the specific problem and 
takes into account the vertex colors in a more judicious manner. 
% when proposing the next state in the chain.
As a result, \algomulti has higher acceptance probabilities than \algofosdick, and mixes faster.
% !TEX root = ../main.tex
\section{Related Work}\label{sec:related}

\iffalse
\mynote[from=Matteo,date=5/30]{Just to put down two papers I found:
  \url{https://arxiv.org/pdf/2402.06466}, and
  \url{https://academic.oup.com/comnet/article-abstract/8/3/cnaa018/5879929}
  (with correction at
  \url{https://academic.oup.com/comnet/article/11/3/cnad014/7172825}), so I
  don't have to keep the tabs open in my browser.
}

\mynote[from=gdfm]{Unpopular proposal: can we do without Related Work?
In the end all the null models are related but none considers labels.
Reviewing the whole literature is impossible, just putting a ref to a couple of surveys might be enough?
\url{https://arxiv.org/html/2403.14415v1}
\url{https://doi.org/10.1007/s10955-018-2039-4}
\url{https://doi.org/10.1145/3369782}
}
\fi

When searching for patterns in network data it is essential to be able to reason about their significance.
In statistics, there is a long tradition of assessing significance by comparing an observed pattern with
its occurrence in a \emph{randomized null model}~\citep{fisher1935design}.
Extending this idea to networks leads to \emph{random-graph null models}, 
where one compares properties observed in real-world networks
with properties observed in networks sampled from a certain random-graph distribution.
While there is a vast literature on random-graph models, 
such as the Erd\H{o}s-R\'{e}nyi random graph~\citep{renyi1959random} and 
the preferential attachment model~\citep{barabasi1999emergence,bollobas2001degree},
which are simple to generate via an iterative sampling process, 
practitioners often seek to sample networks from a space of networks
satisfying certain constraints. 
The most commonly-used constrained random-graph null model is the 
\emph{configuration model}~\citep{bender1978asymptotic,bollobas1980probabilistic,fosdick2017configuring}, 
where the sample space consists of all networks having a specified degree sequence. 
Configuration models have a long research history with applications in 
sociology~\citep{moreno1938statistics}, 
ecology~\citep{connor1979assembly}, 
systems biology~\citep{milo2002network}, and other disciplines.

The \emph{Markov chain Monte Carlo (MCMC)} method is an archetypal approach 
for sampling from the space of networks with a fixed degree sequence. 
The MCMC technique performs a random walk over the sample space \samplespace
of feasible networks and appropriately modifies the transition probabilities of the walk, 
e.g., using the Metropolis-Hastings algorithm~\citep{peskun1973optimum}, 
so that the \emph{stationary distribution} is a desirable target distribution, such as the \emph{uniform} one.
Typically, the neighboring states in the Markov chain are networks that differ only in a  
\emph{two-edge swap}, making it easy to transition between states
and sample random networks from the whole space \samplespace.
By proving that the Markov chain is \emph{strongly connected} (\emph{irreducible}) and \emph{aperiodic}, 
it can be argued that there is a \emph{unique stationary distribution}, 
and the Metropolis-Hastings algorithm can be used to obtain samples from it.

A question of theoretical interest is the \emph{mixing time} of the MCMC method, 
i.e., the number of steps required before the actual sample distribution
is $\epsilon$-close to the stationary distribution. 
Theoretical papers have derived conditions that the method is \emph{rapidly mixing}, 
i.e., the number of required steps is polynomial~\citep{cooper2007sampling,greenhill2014switch}.
However, the general case is still not fully understood. 
Furthermore, existing bounds are high-degree polynomials 
and are mostly of theoretical interest.
In practice, researchers employ various diagnostics to assess empirically the 
MCMC convergence~\citep{roy2019convergence,dutta2021sampling}, such as
comparing the variance of sample graph statistics 
inside a sequence of the chain and against the variance across multiple sequences~\citep{gelman1995bayesian}.

Other types of graph ensembles have been proposed
as a null model to assess the statistical importance of patterns and graph structure. 
Those include 
\emph{maximum-entropy models}~\citep{squartini2017maximum}, 
which however preserve the degree sequence only in expectation,
and \emph{exponential random graph models}~\citep{snijders2002markov},
which increase the probability of observing certain subgraph structures, 
and which are also typically sampled using MCMC methods.

Overall, a large number of methods for sampling graph null models have been presented in the literature, 
% where different graph spaces have been proposed, 
and the ideas have been applied to analyzing data from different disciplines. 
Nevertheless, perhaps surprisingly, no previous work on preserving the color assortativity of a network exists.

% \itodo{We should also say that Polaris-B appears in Fosdick et al.}

% !TEX root = ../main.tex
\section{Preliminaries}\label{sec:prelims}

This section introduces the main concepts and notation used throughout the
work.
We use double curly braces to denote multisets, e.g.,
$A=\mset{a,b,b,\allowbreak c,d,d,d}$,
and $\mul{a}{A}$ is the \emph{multiplicity} of an element $a$ in a multiset~$A$,
i.e., the number of times $a$ appears in $A$. $\card{A}$ denotes the multiset
cardinality of a set, e.g., $\card{A} = 7$ in the example above. The notation
$\setf{A}$ represents the set obtained by removing all duplicates from~$A$,
e.g., $\setf{A} = \set{a,b,c,d}$.

\begin{definition}[Colored Multigraph]\label{def:multigraph}
  A \emph{colored, undirected, multigraph} is a tuple $G \doteq (V, E, \labelset,
  \labelfsym)$, where $V$ is a set of vertices, $E$ is a multiset of edges between
  vertices, each edge being an unordered pair of vertices from $V$, and
  $\labelfsym: V \to \labelset$ is a labeling function assigning a color (i.e., a single label) from
  the set $\labelset$ to each vertex.
  %\rev{The colors in $\labelset$ have equal importance, with no hierarchy among them.}
\end{definition}

We allow (multiple) self-loops, i.e., edges of the type $(u,u)$, $u \in V$. All
multigraphs we consider are colored, so henceforth we use ``multigraph'' to mean ``colored
multigraph''. We refer to edges incident to vertices with the same color as
\emph{monochrome} edges, and edges incident to vertices with different colors as
\emph{bichrome} edges.

Two edges $(u,w),(v,z)$ are \emph{distinct} when they are two
different members of $E$. Two distinct edges may be \emph{copies} of the
same multiedge, i.e., be incident to the same pair of vertices, or to the same
vertex if they are self-loop. We write $(u,w) = (v,z)$ if two edges are
copies, but since edges are unordered pairs of vertices, this notation does not
imply $u = v$ and $w = z$, as it may be $u=z$ and $w=v$.

A multigraph $G=(V,E,\labelset,\labelfsym)$ can be seen as an
integer-weighted graph $G'=(V,E',\labelset,\labelfsym,\mathsf{w})$, with $E'=
\setf{E}$, and $\mathsf{w}$ a function that assigns a natural
weight to edges in $E'$, so that $\mathsf{w}(e) = \mul{e}{E}$.

Given a multigraph $G=(V,E,\labelset,\labelfsym)$, 
for each $u \in V$, $\neighs{u}{G}$ denotes
the multiset of neighbors of $u$, and $\degree{u}{G} \doteq
\card{\neighs{u}{G}}$ the \emph{degree of $u$ in $G$}. 
For each $\ell \in \labelset$,
let $V^\ell \doteq \set{v \in V \suchthat \labelf{v} = \ell}$ be the set of
vertices with color $\ell$.
For each $u \in V$ and $\ell \in \labelset$, let
$\lneighs{u}{G}{\ell} \doteq \mset{v \in V^\ell \suchthat (u,v) \in E}$ be the
multiset of neighbors of $u$ in $G$ with color $\ell$, and let $\nlneighs{u}{G}{\ell}
\doteq \card{\lneighs{u}{G}{\ell}}$.
Clearly, $\degree{u}{G} = \sum_{\ell \in \labelset} \nlneighs{u}{G}{\ell}$.

\begin{definition}[JCM]\label{def:jlm}
  The \emph{Joint Color Matrix} $\jcm{G}$ of a multigraph $G = (V, E, \labelset,
 \labelfsym)$ is the symmetric square matrix $\mathsf{J}_G \in
 \mathbb{N}^{\card{\labelset} \times \card{\labelset}}$ where each entry
 $\jcm{G}[\ell,r]$ is the number of edges between a  vertex with color $\ell$ and a
 vertex with color $r$, i.e.,
 \begin{align*}
   \jcm{G}[\ell,r] \doteq \card{\mset{(u,w) \in E \suchthat \labelf{u} = \ell
   \wedge \labelf{w} = r}}\enspace.
 \end{align*}
\end{definition}

\Cref{fig:lab_graphs} shows two multigraphs with two colors, the same
degree sequence, and the same JCM (shown in the center of the figure).

\subsection{Null Models for Graph Properties}\label{sec:nullmodel}

For any multigraph $G$, let $\props_G$ be a set of properties from $G$,
e.g., the number of edges, the degree sequence, the diameter, or similar
structural properties, which may be scalars, vectors, or matrices.

Let $\observed=(V,E,\labelset,\labelfsym)$ be an observed multigraph. Given
$\props_{\observed}$, the microcanonical \emph{null model} $\nullmodel \doteq (\nullset,
\nullprob)$ is a tuple where $\nullset$ is the set of all and only the
multigraphs $G=(V,E_G,\labelset,\labelfsym)$ on the same set of vertices,
with the same colors and coloring function as $\observed$, and that preserve
each property in $\props_{\observed}$ i.e., such that $\props_G =
\props_{\observed}$, and where $\nullprob$ is a probability distribution over
$\nullset$. Clearly $\observed \in \nullset$, and the graphs in $\nullset$ only
differ by their multisets of edges.

\subsection{Markov Chain Monte Carlo Methods}\label{sec:mcmc}

All algorithms in \algonamecollective follow the \emph{Markov chain Monte Carlo (MCMC)
method}, using the \emph{Metropolis-Hastings (MH) approach} \citep[Ch.\ 7 and
10]{MitzenmacherU05}. Let us now introduce these concepts.

Let $\mathcal{G} = (\states, \mathcal{E}, \mathsf{w})$ be a directed,
weighted, strongly connected, and aperiodic\footnote{A graph is aperiodic iff
the greatest common divisor of the lengths of its cycles is 1.} graph, which
may have self-loops. The vertices in $\states$ are known as \emph{states}, and
$\mathcal{G}$ is known as a state graph. Using the same notation we defined
previously, for any state $s \in \states$, $\neighs{s}{\mathcal{G}}$ denotes the
set of (out-)\emph{neighbors} of $s$, i.e., the set of states $u$ s.t.\ $(s,u)
\in \mathcal{E}$. Iff $u \in \neighs{s}{\mathcal{G}}$, then $\mathsf{w}(s,u) >
0$, and it holds $\sum_{u \in \neighs{s}{\mathcal{G}}} \mathsf{w}(s,u) = 1$.
Thus, for any $u \in \states$, we can define the \emph{transition probability
$\transprob{s}{u}$ from $s$ to $u$} as $\mathsf{w}(s,u)$ if $u \in
\neighs{s}{\mathcal{G}}$, and 0 otherwise.

Given any $\mathcal{G}$ as above, a \emph{neighbor proposal probability
distribution $\neighprobsym{v}$} over $\neighs{v}{\mathcal{G}}$ for any $v \in
\states$, and any probability distribution $\statprob$
over $\states$, the \emph{MH approach} is a generic
procedure to sample a state $s \in \states$ according to $\statprob$. Starting
from any $v \in \states$, one first draws a neighbor $u$ of $v$ according to
$\neighprobsym{v}$, and then ``moves'' to $u$ with probability
\begin{align*}
  \acceptprob{v}{u} = \min \left\{ 1, \frac{\statprob(u)}{\statprob(v)}
  \frac{\neighprob{u}{v}}{\neighprob{v}{u}} \right\},
\end{align*}
otherwise remains in $v$. The quantity $\acceptprob{v}{u}$ is known as the
\emph{acceptance probability} of $u$. The sequence of states obtained by
repeating this procedure forms a Markov chain over $\states$ with unique
stationary distribution $\statprob$. Thus, after a sufficiently large number of
steps $t$, the state $v_t$ at time $t$ is %(either approximately or exactly)
distributed according to $\statprob$, and can be considered a sample of
$\states$ according to $\statprob$.

To use MH, it is necessary to define: \textit{(i)} the graph
$\mathcal{G}$ as above, taking special care in ensuring that it is
strongly-connected and aperiodic; \textit{(ii)}, the neighbor sampling
distribution $\neighprob{s}$ for every state $s \in \states$; and \textit{(iii)}
the desired sampling distribution $\statprob$ over $\states$.

% !TEX root = ../main.tex
\section{A Null Model for Vertex-Colored Graphs}\label{sec:jlmnullmodel}

Given an observed $\observed \doteq (V,E,\labelset,\labelfsym)$, with $V=\{v_1, \dotsc,
v_{\card{V}}\}$, we consider the null model $\nullmodel = (\nullset,
\nullprob)$ where $\props_{\observed}$ consists of the degree sequence $\left[d_{\observed}(v_1), \dotsc, d_{\observed}\left(v_{\card{V}}\right)\right]$ and the JCM $\jcm{\observed}$.

This null model is essentially the simplest one that considers
the color information, if one assumes that the color of a vertex is an intrinsic
property.
While one could think of preserving only the colored degree sequences for
each color, doing so is equivalent to preserving the ``generic'' degree
sequence, and thus does not leverage the color information in any meaningful way.
Indeed, this can be done on the unlabeled version of the multigraph~\citep{fosdick2018configuring}.

Our goal is to design efficient MCMC algorithms to sample from $\nullset$
w.r.t.\ $\nullprob$ as defined above. We first define two operations that allow
transforming a multigraph $G$ into a multigraph $H$, potentially identical to
$G$. The first operation is the classic Double Edge Swap (DES), known under many
names and introduced many times in the
literature~\citep{cafieri2010loops,artzy2005generating,verhelst2008efficient,taylor2006contrained,stone1990checkerboard,gotelli1996null}.

\begin{definition}[Double Edge Swap (DES)]\label{def:des}
  Given a multigraph $G \doteq (V, E, \labelset, \labelfsym)$, let $(u,w),
  (v,z)$ be two distinct edges in $E$. Consider the multigraph $H = (V, (E
  \setminus \mset{(u,w), (v,z)}) \cup \mset{(u,z), (w,v)}, \labelset,
  \labelfsym)$. We
  call the operation that ``swaps'' $(u,w), (v,z)$ with $(u,z), (w,v)$ a
  \emph{Double Edge Swap (DES)}, and denote it
  $\swap{(u,w),(v,z)}{(u,z),(w,v)}$.
\end{definition}

We say that a DES is \emph{applied to} the origin multigraph $G$ to obtain the
destination multigraph $H$, or that a DES \emph{transforms $G$ into $H$}.

For every unordered pair $((u,w), (v,z))$ of distinct edges in the origin graph,
there are exactly two DESs that involve them: $\swap{(u,w),(v,z)}{(u,z),(v,w)}$ and
$\swap{(u,w),(v,z)}{(u,v),(z,w)}$. If the destination multigraph $H$ is the same
for both DESs, we say that the DESs are \emph{equivalent}. If $H = G$, we say
that the DES is a \emph{no-op}, otherwise we say that the DES is a \emph{moving}
DES\@. For the same unordered pair of distinct edges in the origin graph, one
DES may be a no-op, and the other may be a moving DES.%

Multiple expressions may correspond to the same DES, as a DES is defined by the
multiset of edges in the origin multigraph and by the multiset of edges in the
destination multigraph. For example, the expressions
$\swap{(u,w),(v,z)}{(u,z),(w,v)}$ and $\swap{(z,v),(u,w)}{(u,z),(w,v)}$ both
denote the same DES\@.

DESs can be used in MCMC algorithms to sample from a null model that preserves
the degree sequence of an observed multigraph~\citep[and references
therein]{fosdick2018configuring}: given a DES, the destination multigraph has
the same vertices and the same degree sequence as the origin.
Conversely, the JCM may or may not be preserved by a DES.
Thus we define the following specific operation.

\begin{definition}[JCM-preserving Double Edge Swap (JDES)]\label{def:jdes}
  A \emph{JCM-preserving Double Edge Swap (JDES)} is a DES such that the
  destination multigraph $H$ retains the JCM of the origin multigraph~$G$.
\end{definition}

An example of JDES that can be applied to the left multigraph in
\Cref{fig:lab_graphs} is $\swap{(1,7), (3,6)}{(1,6),(3,7)}$, while the operation $\swap{(3,4),
(9,8)}{(3,8), (9,4)}$ is a DES but not a JDES.%

%\inote{Maybe it would be nice to show one that goes from the left graph to the
% right graph?}
%\mynote[from=Giulia]{The two graphs are not neighbors, and to go from the left
%  to the right graph we need to perform swaps that add edges not present in the
%  right graph. An example of sequence of LSOs that connects the two graphs is:
%  \begin{itemize}
%      \item $\lso{(3,4), (9,8)}{(3,8), (9,4)}$;
%      \item $\lso{(1,7), (8,3)}{(1,3), (8,7)}$;
%      \item $\lso{(1,2), (7,5)}{(1,5), (7,2)}$;
%      \item $\lso{(1,5), (4,6)}{(1,6), (4,5)}$.
%  \end{itemize}
%}

For any unordered pair $((u,w), (v,z))$ of distinct edges in the origin
multigraph, zero, one, or both DESs may be JDESs. We
now give a complete characterization of which DESs are JDESs, considering
different cases based on the properties of the edges involved.

% \begin{description}
  % \item[Case 0:] 
  \spara{Case 0:} $\{\labelf{u}, \labelf{w}\} \cap \{ \labelf{v}, \labelf{z} \}
  = \varnothing$, i.e., the two edges have disjoint vertex colors. Then, neither
  DES is a JDES.%

  % \item[Case 1:]
  \spara{Case 1:} $\card{\{u,w,v,z\}} = 1$, i.e., $(u,w)$ and $(v,z)$ are copies
    of the same self-loop multiedge. Both DESs are JDESs and no-ops.

  % \item[Case 2A:] 
  \spara{Case 2A:} $\card{\{u,w,v,z\}} = 2 \wedge u=w \wedge v=z \wedge
    \labelf{u} = \labelf{v}$, i.e., $(u,w)$ and $(v,z)$ are two different self-loops on vertices with the same color. Both DESs are equivalent 
    JDESs and moving DESs.

  % \item[Case 2B:] 
  \spara{Case 2B:} $\card{\{u,w,v,z\}} = 2 \wedge u \neq w \wedge v \neq z \wedge \labelf{u} = \labelf{w}$, i.e., $(u,w)$ and $(v,z)$ are identical
    non-self-loop monochrome multiedges. Both DESs are JDESs; one is
    a no-op, while the other creates a self-loop.% on $u$ and on $w$.

  % \item[Case 2C:] 
  \spara{Case 2C:} $\card{\{u,w,v,z\}} = 2 \wedge u \neq w \wedge v \neq z \wedge \labelf{u} \neq \labelf{w}$, i.e., $(u,w)$ and $(v,z)$ are identical non-self-loop bichrome multiedges. Only one DES is a JDES, and is a no-op.

  % \item[Case 2D:] 
  \spara{Case 2D:} $\card{\{u,w,v,z\}} = 2 \wedge (u = w \vee v = z) \wedge \neg
    (u = w \wedge v = z)$, i.e., one edge is a self-loop, the other is not but is incident to the self-loop vertex. Both DESs are JDESs and no-ops.

  % \item[Case 3A:] 
  \spara{Case 3A:} $\card{\{u,w,v,z\}} = 3 \wedge (w=u \vee v=z) \wedge
    \{\labelf{u}, \labelf{w}\} \cap \{ \labelf{v}, \labelf{z} \} \neq
    \varnothing$, i.e., one edge is a self-loop, the other is not and is incident to different vertices than the self-loop, and at least one of these vertices has the same color as the self-loop. Both DESs are equivalent JDESs and moving DESs.

  % \item[Case 3B:] 
  \spara{Case 3B:} $\card{\{u,w,v,z\}} = 3 \wedge u \neq w$ $\wedge v \neq z$ $\wedge \card{\{\labelf{u}, \labelf{w}, \labelf{v}, \labelf{z}\}}$ $=$ $1$, i.e.,
    neither edge is a self loop, and since $\card{\{u,w,v,z\}} = 3$, it holds
    exactly one of $u = v$, $u = z$, $w = v$, or $w = z$, so the edges form a
    wedge with vertices sharing the same color. Assume, w.l.o.g., that $u = v$. Then both 
    DESs are JDESs; one is a no-op, while the other creates a self-loop on
    $u$ and an edge between the vertices at the extremes of the former wedge.

  % \item[Case 3C:] 
  \spara{Case 3C:} $\card{\{u,w,v,z\}} = 3 \wedge u \neq w \wedge v \neq z \wedge
    \card{\{\labelf{u}, \labelf{w}, \labelf{v}, \labelf{z}\}}$ $=$ $2$ $\wedge$ $(\labelf{u} = \labelf{w} \vee \labelf{v} = \labelf{z})$. Similar to Case 3B, but exactly one edge is monochrome. Both DESs are JDESs; one is a no-op, while the other creates a self-loop on $u$ and an edge between the two extremes of the former wedge.

  % \item[Case 3D:] 
  \spara{Case 3D:} $\card{\{u,w,v,z\}} = 3 \wedge u \neq w \wedge v \neq z \wedge \card{\{\labelf{u}, \labelf{w}, \labelf{v}, \labelf{z}\}}$ $=$ $2$ $\wedge$ $\labelf{u} \neq \labelf{w} \wedge \labelf{v} \neq \labelf{z}$. The edges form a wedge where the endpoints of the wedge have the same color and the vertex in the middle has a different color. One DES is a JDES and is a no-op.

  % \item[Case 3E:] 
  \spara{Case 3E:} $\card{\{u,w,v,z\}} = 3 \wedge u \neq w \wedge v \neq z \wedge \card{\{\labelf{u}, \labelf{w}, \labelf{v}, \labelf{z}\}}$ $=$ $3$. The edges form a wedge, but all three vertices have different colors. One DES is a JDES and is a no-op.

  % \item[Case 4A:] 
  \spara{Case 4A:} $\card{\{u,w,v,z\}} = 4 \wedge \card{\{\labelf{u}, \labelf{w}, \labelf{v}, \labelf{z} \}}$ $=$ $3$ $\wedge$ $\labelf{u} \neq \labelf{w} \wedge \labelf{v} \neq \labelf{z}$. The edges are incident to four
    distinct vertices, neither of them is monochrome, they are not both bichrome
    with the same two colors, but they are incident to one vertex with the same
    color. One DES is a JDES and is a moving DES.%

  % \item[Case 4B:] 
  \spara{Case 4B:} $\card{\{u,w,v,z\}} = 4 \wedge \card{\{\labelf{u}, \labelf{w}, \labelf{v}, \labelf{z} \}}$ $=$ $2$ $\wedge$ $\labelf{u} \neq \labelf{w} \wedge \labelf{v} \neq \labelf{z}$. The edges are incident to four
    distinct vertices and are both bichrome with the same two colors.
    One DES is a JDES and is a moving DES.%

  % \item[Case 4C:] 
  \spara{Case 4C:}
  $\card{\{u,w,v,z\}} = 4 \wedge \card{\{\labelf{u}, \labelf{w},
    \labelf{v}, \labelf{z} \}} \in \{1,2\}$ $\wedge$ $(\labelf{u} = \labelf{w} \vee
    \labelf{v} = \labelf{z})$. The edges are incident to four
    distinct vertices, with at least three of them having the same color. Both DESs are JDESs, they are not equivalent, and both are moving DESs.
% \end{description}

There cannot be more than two JDESs transforming $G$ into $H$, for $H \neq G$.
If there are two, they involve the same pair of edges and are
equivalent. All JDESs are reversible: if there are JDESs transforming $G$
into $H$, then there are JDESs transforming $H$ into $G$.

\subsection{Strong Connectivity and Aperiodicity of
%\subsection{Ergodicity of
  \texorpdfstring{$\nullset$}{Z} via JDESs}\label{sec:strongconnect}

In \algoname, the state space $\mathcal{S}$ of the state
graph $\mathcal{G} = (\mathcal{S}, \mathcal{E}, \mathsf{w})$ is $\nullset$, and
the desired probability distribution according to which to sample is
$\nullprob$. The edge set $\mathcal{E}$ is defined as follows. For any $G, H \in
\nullset$, there is an edge $(G,H) \in \mathcal{E}$ if there is a JDES
from $G \in \nullset$ to $H \in \nullset$. Clearly, if that is the case, there
is also an edge $(H,G) \in \mathcal{E}$ as all JDESs are reversible.
Additionally, there may be a self-loop in
$G$ even if there is no JDES from $G$ to $G$, but $G$ has a neighbor $H$ such
that the acceptance probability $\acceptprob{G}{H}$ is strictly less than 1.

We say that $G$ and $H$ are \emph{neighbors} iff there is a JDES transforming
$G$ into $H$. As required by MH, we show that the resulting state graph is
strongly connected (\Cref{thm:connect}) and aperiodic (\Cref{thm:aperiod})
(full proofs in \Cref{ax:proofs}), which ensures the chain is ergodic.

\begin{theorem}\label{thm:connect}
  The state space $\nullset$ is strongly connected by JDESs.
\end{theorem}

The proof of this theorem explicitly builds a sequence of JDESs from any
state $G \in \nullset$ to any other $H \in \nullset$, by first going from $G$ to
a $\tilde{G} \in \nullset$ such that every vertex has in $\tilde{G}$
exactly the same number of neighbors of each color as it has in $H$, then
going from $\tilde{G}$ to $H$.

\begin{theorem}\label{thm:aperiod}
  Given a multigraph $G$, if either of the following conditions holds, then the state graph $\mathcal{G}$ is aperiodic:
  \begin{itemize}
    \item there exist two edges $(u,w)$ and $(v,z)$ that fall in cases 1, 2A,
      2B, 2C, 2D, 3A, 3B, 3C, 3D, 3E, or 4C of the
      classification; or
    \item there exist a color $\ell \in \labelset$ such that there are bichrome
      edges $(u,v)$, $(u,z)$, $(w,x)$, with all of $u$, $v$, $z$, $w$ and $x$
      distinct, and $\labelf{u} = \labelf{w} = \ell$.
  \end{itemize}
\end{theorem}

The proof of \Cref{thm:aperiod} involves a case-by-case analysis of the JDES,
showing that either there must be a self-loop on a vertex in $\mathcal{G}$, or
there are cycles of length 3 and 2, thus ensuring aperiodicity.

The conditions in \Cref{thm:aperiod} are extremely mild. For example, the first
condition implies that if there is a color such that there are two monochrome
edges with that color, then the state graph is aperiodic. Only a (relatively)
small class of unusual multigraphs results in periodic state graphs.
Additionally, the conditions are not necessary for the graph to be aperiodic:
the algorithms we present run Markov chains on a state
graph that may have additional self-loops, as they are based on MH.
%, due to the fact that a state $G$ may
%have a neighbor $H$ for which the acceptance probability $\acceptprob{G}{H}$ is
%less than 1, i.e., a proposed move may be rejected with some probability,
%keeping the chain in the same state.\iwarning{We kinda say this twice, also
%before the theorems. Maybe keep it once only?}

\subsection{A first baseline algorithm}\label{sec:algofosdick}

To warm up, we present a baseline algorithm \algofosdick, which is an adaptation
of \citet[Algorithm 3]{fosdick2017configuring}\footnote{This algorithm only appears in the arXiv version of this paper~\citep{fosdick2018configuring}.}
to our task of interest. \Citet{fosdick2017configuring} introduced the algorithm
to sample uniformly from the space of unlabeled multigraphs with the same degree
sequence.\@ \algofosdick is a tailored version of this algorithm to sample
according to any distribution $\nullprob$ from the space of colored multigraphs
with the same degree sequence and the same JCM\@.

\algofosdick (pseudocode in \Cref{alg:fosdick}) starts by setting the current
state $G$ of the Markov chain to the observed multigraph $\observed$
(\Cref{algline:initgf}). It then enters a loop for $t$ iterations. At each
iteration, it first samples two edges $e_1$ and $e_2$ uniformly at random from
the population of ordered pairs of distinct edges
(Lines~\ref{algline:edge1samplef}--\ref{algline:edge2samplef}). The algorithm
then randomly chooses one of the two possible DESs involving $e_1$ and $e_2$
(\Cref{algline:des}). If the selected DES $\mathsf{des}$ is not a JDES
(\Cref{algline:notjdes}), the algorithm samples a new DES; if it is a no-op (\Cref{algline:noopf}), the Markov
chain stays in $G$. Otherwise, the algorithm computes a value $\rho$ that
depends on properties of the sampled edges, which is used as follows to ensure
that the stationary distribution of the Markov chain is $\nullprob$. Let $H \neq
G$ be the multigraph obtained by applying the JDES `$\mathsf{des}$' to $G$
(\Cref{algline:neighborf}).\@ \algofosdick checks if $\rho
\nullprob(H)/\nullprob(G)$ is greater than a real number sampled uniformly at random
from $[0,1]$, and if so, it sets the current state $G$ of the Markov chain to
$H$ (\Cref{algline:acceptancef}), otherwise the chain remains in $G$.

The only difference in \algofosdick w.r.t.\ \citep[Algorithm
3]{fosdick2017configuring} is that it checks if the sampled DES is a JDES, and keeps sampling a new DES until it is a JDES (\Cref{algline:notjdes}).

\begin{theorem}\label{thm:algofosdickcorrect}
  The Markov chain run by \algofosdick has stationary distribution $\nullprob$.
\end{theorem}

The complete proof is in \Cref{ax:proofs}, and shows that \algofosdick
 follows the MH approach, thus ensuring the thesis.

In practice, the number of steps $t$ must be chosen in such a way that the multigraph
returned by \algofosdick is, at least approximately, distributed according to
$\nullprob$, i.e., $t$ should be greater or equal to the mixing time for the
Markov chain. Theoretical results for the mixing time of these Markov chains are hard to obtain:
even in the case of the state space of multigraphs connected
by DESs (i.e., when only the degree sequence is preserved), upper bounds on the
mixing time are only known in limited cases \citep{erdHos2022mixing}.
Therefore, \Cref{sec:exper} presents an empirical evaluation of the behavior of $t$.

\begin{algorithm}[t]
  \footnotesize
  \caption{\algofosdick}\label{alg:fosdick}
  \DontPrintSemicolon%
  \SetKwFor{RepTimes}{repeat}{times}{end}
  \SetKwRepeat{Do}{do}{while}
  \SetKw{Continue}{continue}
  \SetKw{Or}{or}
  \SetKw{And}{and}
  \KwIn{Observed multigraph $\observed \doteq (V, E, \labelset,\labelfsym)$,
    distribution $\nullprob$ over $\nullset$, number of iterations $t$}
  \KwOut{Multigraph drawn from $\nullset$ according to $\nullprob$}
  $G \gets \observed$\label{algline:initgf}\;
  \RepTimes{$t$}{
    \Do{$\mathsf{des}$ is not a JDES \emph{ // Case 0}\label{algline:notjdes}}{
      $e_1=(u,w) \gets$ edge drawn u.a.r.\ from $E$\;\label{algline:edge1samplef}
      $e_2=(v,z) \gets$ edge drawn u.a.r.\ from $E \setminus
      \{e_1\}$\;\label{algline:edge2samplef}
      $\mathsf{des} = (\swap{e_1,e_2}{e_1',e_2'}) \gets$ DES drawn u.a.r.\ from
      $\{\swap{(u,w),(v,z)}{(u,z),(v,w)},\ \swap{(u,w),(v,z)}{(u,v),
      (w,z)}\}$\;\label{algline:des}
    }
    % \lIf(// Case 0){$\mathsf{des}$ is not a JDES}{%
      % \Continue\label{algline:notjdes}
    % }
    \lIf(// Cases 1, 2C, 2D, 3D, 3E, and no-ops for other cases){$\mathsf{des}$ is a no-op}{%
      \Continue\label{algline:noopf}
    }
    \If(// Cases 4\lparen%
         A,B,C\rparen, moving DES){$\card{\{u, w, v, z\}} = 4$}{
      $\rho \gets \large\sfrac{(\mul{e_1'}{G}+1)(\mul{e_2'}{G}+1)}{\mul{e_1}{G}\mul{e_2}{G}}$\;
    }
    \ElseIf{$\card{\{u, w, v, z\}} = 3$}{
      \If(// Case 3A){$e_1$ is a self-loop \Or\ $e_2$ is a self-loop}{
        $\rho \gets \large\sfrac{(\mul{e_1'}{G}+1)(\mul{e_2'}{G}+1)}{2\mul{e_1}{G}\mul{e_2}{G}}$\;
      }
      \Else(// Cases 3B and 3C, moving DES){
        $\rho \gets \large\sfrac{2(\mul{e_1'}{G}+1)(\mul{e_2'}{G}+1)}{\mul{e_1}{G}\mul{e_2}{G}}$\;
      }
    }
    \Else(// i.e., $\card{\{u, w, v, z\}} = 2$){
      % \mynote[from=Matteo,date=6/6]{
      %   \Citep[Algorithm 3]{fosdick2017configuring} has a first subcase here for
      %   ``exactly one of $e_1$ and $e_2$ is a self-loop``. But since it holds
      %   $\card{\{u, w, v, z\}} = 2$, this would fall in case 2D, where both
      %   JDESs are no-ops, so it was handled on \Cref{algline:noopf}.}
      %\If(// Case 2D){exactly one of $e_1$ and $e_2$ is a self-loop}{
      %    $\rho \gets \large\sfrac{(\mul{e_1'}{G}+1)(\mul{e_2'}{G}+1)}{2\mul{e_1}{G}\mul{e_2}{G}}$\;
      %}
      %ElseIf
      \If(// Case 2A){both $e_1$ and $e_2$ are self-loops}{
          $\rho \gets \large\sfrac{(\mul{e_1'}{G}+2)(\mul{e_1'}{G} + 1)}{4\mul{e_1}{G}\mul{e_2}{G}}$
      }
      \Else(// Case 2B){
          $\rho \gets \large\sfrac{4(\mul{e_1'}{G}+1)(\mul{e_2'}{G}+1)}{\mul{e_1}{G}(\mul{e_1}{G}-1)}$
      }
    }
    $H \gets $ apply $\mathsf{des}$ to $G$\;\label{algline:neighborf}
    \lIf{$\mathsf{Uniform}(0,1) < \Large\sfrac{\rho\nullprob(H)}{\nullprob(G)}$}{%
      $G \gets H$\label{algline:acceptancef}
    }
  }
  \Return{$G$}\;
\end{algorithm}

\subsection{An algorithm tailored to the task}\label{sec:algomulti}

We now present \algomulti, a color-aware algorithm to
sample a multigraph from $\nullset$
according to $\nullprob$ by leveraging the properties of the data and of the
task better than \algofosdick (\Cref{sec:algofosdick}). As the results of our
experimental evaluation show (\Cref{sec:exper}), this algorithm has higher
acceptance probability and converges faster than the baseline presented earlier.

\algomulti improves over \algofosdick in several ways:
\begin{squishlist}
  \item it avoids sampling pairs of distinct edges falling in Case 0 of the
    characterization of JDES, i.e., pairs of distinct edges such that neither DES
    involving them is a JDES;
  \item if the sampled pair of distinct edges is such that one of the DESs
    is a no-op or not a JDES and the other is a moving JDES
   (Cases 2B, 2C, 3B, 3C) \algomulti always chooses the moving one;
  \item if the sampled pair of distinct edges is such that the JDESs involving
    them are equivalent (Cases 1, 2A, 2D, 3A), it deterministically chooses one
    thus avoiding random choices.
\end{squishlist}

As \algomulti avoids selecting no-op JDESs in some cases (second bullet
point above), one should ask whether the resulting state graph is still
aperiodic under the conditions stated in \Cref{thm:aperiod}. The answer is that
the first condition should be modified to hold only for Cases 1, 2C, 2D, 3D, and
4C. The condition for 4C is particularly mild: it only requires the existence
of a color $\ell \in \labelset$ such that there are two non-self-loop,
non-copies, monochrome edges with color $\ell$, i.e., two edges involving four
distinct vertices with the same color.

All the above improvements of \algomulti over \algofosdick reduce the
probability that the Markov chain remains in the current state, while
not decreasing (and potentially increasing) the transition probability from a
state to any of its different neighbors. In other words, the off-diagonal entries
of the transition matrix of the Markov chain realized by \algomulti are not
smaller than the corresponding entries in the transition matrix of the Markov
chain realized by \algofosdick. \algomulti therefore precedes \algofosdick in
Peskun's order~\citep{peskun1973optimum}, which implies that it has a smaller
mixing time, i.e., requires fewer steps for the state of the chain to be
(approximately) distributed according to the stationary distribution.

\algomulti takes into account the number of \emph{different} colors $\mathsf{nl}(A)$
in a multiset of vertices $A$ to distinguish various cases of the JDES.
Formally, w.l.o.g, let $\labelset = \{0, \dotsc, k-1\}$, for some $k > 1$. For any
unordered $(\ell, r) \in \labelset \times \labelset$, let
$E_{G,\ell,r}$ be the multiset of edges incident to one vertex with color $\ell$ and
one vertex with color $r$. Clearly $E_{G,\ell,\ell}$ is the multiset of the
monochrome edges with color $\ell$. We also define
\[
  E_{G,\ell} \doteq \bigcup_{r \in \labelset} E_{G,\ell,r}
\]
as the multiset of edges incident to at least one vertex with color $\ell$.
Given a multiset $A$ of vertices, let
\[
  \mathsf{nl}(A) \doteq \card{\{\lambda(v) \suchthat v \in A\}}.
\]

\Cref{alg:algomultiset} presents \algomulti's pseudocode. The algorithm takes as
input the observed multigraph $\observed$, the distribution $\nullprob$
according to which one wants to sample from $\nullset$, and a number $t$ of
iterations. It keeps track of the current state of a Markov chain on $\nullset$
in a variable $G$ initialized to $\observed$ (\Cref{algline:initg}).
\algomulti then enters a loop for $t$ iterations. At each iteration, it first
samples a color $\ell$ from $\labelset$ uniformly at random
(\Cref{algline:labsample}), then it draws two distinct edges $(u,w)$ and $(v,z)$
uniformly at random respectively from $E_{G,\ell}$ and from $E_{G,\ell}
\setminus \{(u,w)\}$
(Lines~\ref{algline:edge1sample}--\ref{algline:edge2sample}). It then checks
which case of the JDES characterization should be considered.
It either sets the variable $\mathsf{jdes}$ to be a moving JDES involving $(u,w)$ and
$(v,z)$, possibly by choosing it uniformly at random when there are two
non-equivalent, moving, JDESs (only happens in Case 4C,
Lines~\ref{algline:case4cstart}--\ref{algline:case4cend}), or keeps the state of
the Markov chain to be the current multigraph $G$ if the JDESs in the considered
case are both no-ops (Cases 2C, 2D, 3D). In the cases when $\mathsf{jdes}$ is
set, the algorithm also sets the variable $\rho$ to a value that, as we discuss
in the analysis of \algomulti (\Cref{thm:algomulticorrect}), ensures that the
multigraph returned by the algorithm is drawn from $\nullset$ according to
$\nullprob$. Let now $H$ be the multigraph obtained by applying $\mathsf{jdes}$
to $G$.\@ \algomulti checks whether a value chosen uniformly at random in
$[0,1]$ is smaller than $\rho\nullprob(H)/\nullprob(G)$, and if so, updates the
state $G$ of the Markov chain to $H$ (\Cref{algline:acceptance}), otherwise the
chain remains in the current state. After $t$ iterations, the current state $G$
is returned.

\begin{theorem}\label{thm:algomulticorrect}
  The Markov chain run by \algomulti has stationary distribution $\nullprob$.
\end{theorem}

The proof can be found in \Cref{ax:proofs}. It essentially shows that, no
matter into what case of the classification the sampled JDES falls, the
algorithm follows the MH approach for choosing the acceptance probability to
ensure the thesis of the theorem.

\begin{algorithm}[htbp]
  \footnotesize
  \caption{\algomulti}\label{alg:algomultiset}
  \DontPrintSemicolon%
  \SetKwFor{RepTimes}{repeat}{times}{end}
  \SetKwFunction{FairRandomCoinFlip}{fairCoinFlip}
  \SetKw{Continue}{continue}
  \SetKw{Or}{or}
  \SetKw{And}{and}
  \KwIn{Observed multigraph $\observed \doteq (V, E, \labelset,\labelfsym)$, distribution
    $\nullprob$ over $\nullset$, number of iterations $t$}
  \KwOut{Multigraph drawn from $\nullset$ according to $\nullprob$}
  $G \gets \observed$\label{algline:initg}\;
  \RepTimes{$t$}{\label{algline:loopstart}
    $\ell \gets$ color drawn u.a.r.\ from $\labelset$\;\label{algline:labsample}
    $(u,w) \gets$ edge drawn u.a.r.\ from
    $E_{G,\ell}$\label{algline:edge1sample}\;
    $(v,z) \gets$ edge drawn u.a.r.\ from $E_{G,\ell} \setminus
    \{(u,w)\}$\label{algline:edge2sample}\;
    \lIf(// Case 1){$\card{\{u,w,v,z\}} = 1$}{\label{algline:case1}
     \Continue% 
    }
    \ElseIf{$\card{\{u,w,v,z\}} = 2$}{
      \If(// Case 2A){both $(u,w)$ \And\ $(v,z)$ are self-loops}{
        $\mathsf{jdes} \gets \swap{(u,u), (v,v)}{(u,v),(v,u)}$\;
        $\rho \gets \large\sfrac{(\mul{(u,v)}{G} +
        2)(\mul{(u,v)}{G}+1)}{\mul{(u,u)}{G}\mul{(v,v)}{G}}$\;\label{algline:rho2A}
      }
      \ElseIf(// Case 2B){neither $(u,w)$ nor $(v,z)$ is a self-loop \And\ $\labelf{u} = \labelf{w}$}{
          W.l.o.g.\ let $u = z$ (thus $w = v$)\;
        $\mathsf{jdes} \gets \swap{(u,v), (v,u)}{(u,u),(v,v)}$\;
        $\rho \gets \large\sfrac{(\mul{(u,u)}{G} +
        1)(\mul{(v,v)}{G}+1)}{\mul{(u,v)}{G}(\mul{(u,v)}{G}-1)}$\;\label{algline:rho2B}
      }
      \lElse(// Case 2C or 2D){\label{algline:continue2C}
        \Continue%
      }
    }
    \ElseIf{$\card{\{u,w,v,z\}} = 3$}{
      \If(// Case 3A){either $(u,w)$ or $(v,z)$ is a self-loop}{
        W.l.o.g.\ let $(u,w)$ be the self-loop\;
        $\mathsf{jdes} \gets \swap{(u,u),(z,v)}{(u,v),(z,u)}$\;
        $\rho \gets \large\sfrac{(\mul{(u,v)}{G} +
        1)(\mul{(u,z)}{G}+1)}{\mul{(u,u)}{G}\mul{(v,z)}{G}}$\;\label{algline:rho3A}
      }
      \Else(// W.l.o.g.\ assume $u=v$){
        \If(// Case 3B){$\mathsf{nl}(u,w,v,z) = 1$}{
          $\mathsf{jdes} \gets \swap{(u,w),(z,u)}{(u,u),(z,w)}$\;
          $\rho \gets \large\sfrac{(\mul{(u,u)}{G} +
          1)(\mul{(w,z)}{G}+1)}{\mul{(u,w)}{G}\mul{(u,z)}{G}}$\;\label{algline:rho3B}
        }
        \ElseIf(// Case 3C){$\labelf{u} = \labelf{w}$ \Or\ $\labelf{v} = \labelf{z}$}{
          $\mathsf{jdes} \gets \swap{(u,w),(z,u)}{(u,u),(z,w)}$\;
          $\rho \gets \large\sfrac{(\mul{(u,u)}{G} +
          1)(\mul{(w,z)}{G}+1)}{\mul{(u,w)}{G}\mul{(u,z)}{G}}$\;\label{algline:rho3C}
        }
        \lElse(// Case 3D or 3E){\label{algline:continue3D}
          \Continue%
        }
      }
    }
    \Else(// i.e., $\card{\{u,w,v,z\}} = 4$){
      \If(// Case 4A){$\mathsf{nl}(u,w,v,z) = 3$ \And\ $\labelf{u} \neq
        \labelf{w}$ \And\ $\labelf{v} \neq \labelf{z}$}{
        W.l.o.g.\ let $\labelf{u} = \labelf{v}$\;
        $\mathsf{jdes} \gets \swap{(u,w),(v,z)}{(u,z),(v,w)}$\;
        $\rho \gets \large\sfrac{(\mul{(u,z)}{G} +
        1)(\mul{(v,w)}{G}+1)}{\mul{(u,w)}{G}\mul{(v,z)}{G}}$\;\label{algline:rho4A}
      }
      \ElseIf(// Case 4B){$\mathsf{nl}(u,w,v,z) = 2$ \And\ $\labelf{u} \neq
        \labelf{w}$ \And\ $\labelf{v} \neq \labelf{z}$}{
        W.l.o.g.\ assume $\ell = \labelf{u} = \labelf{v}$ and let $\ell' \doteq
        \labelf{w} = \labelf{z}$ (it holds $\ell \neq \ell'$)\;
        $\mathsf{jdes} \gets \swap{(u,w),(v,z)}{(u,z),(v,w)}$\;
        $\rho \gets \frac{\frac{\left(\mul{(u,z)}{G}+1\right)\left(\mul{(v,w)}{G}+1\right)}
        {\card{E_{G,\ell}} \left(\card{E_{G,\ell}}-1\right)} + 
        \frac{\left(\mul{(u,z)}{G}+1\right)\left(\mul{(v,w)}{G}+1\right)}{\card{E_{G,\ell'}}\left(\card{E_{G,\ell'}} - 1\right)}}
        {\frac{\mul{(u,w)}{G}\mul{(v,z)}{G}}{\card{E_{G,\ell}}\left(\card{E_{G,\ell}} - 1\right)} + 
        \frac{\mul{(u,w)}{G}\mul{(v,z)}{G}}{\card{E_{G,\ell'}}\left(\card{E_{G,\ell'}}-1\right)}}$\;\label{algline:rho4B}
    }
      \Else(// Case 4C){
          \If{\FairRandomCoinFlip{} is head}{\label{algline:case4cstart}
          $\mathsf{jdes} \gets \swap{(u,w),(v,z)}{(u,z),(v,w)}$\;
          $\rho \gets \large\sfrac{(\mul{(u,z)}{G} +
          1)(\mul{(v,w)}{G}+1)}{\mul{(u,w)}{G}\mul{(v,z)}{G}}$\;\label{algline:rho4C1}
        }
        \Else{%
          $\mathsf{jdes} \gets
          \swap{(u,w),(z,v)}{(u,v),(z,w)}$\;\label{algline:case4cend}
          $\rho \gets \large\sfrac{(\mul{(u,v)}{G} +
          1)(\mul{(z,w)}{G}+1)}{\mul{(u,w)}{G}\mul{(v,z)}{G}}$\;\label{algline:rho4C2}
        }
      }
    }
    $H \gets$ apply $\mathsf{jdes}$ to $G$\;
    \lIf{$\mathsf{Uniform}(0,1) < \Large\sfrac{\rho \nullprob(H)}{\nullprob(G)}$}{%
      $G \gets H$\label{algline:acceptance}
    }
  }
  \Return{$G$}\;
\end{algorithm}

% !TEX root = ../main.tex
\section{Experimental evaluation}\label{sec:exper}
Our experimental evaluation has three objectives. 
First, we demonstrate the qualitative differences between multigraphs sampled using the traditional configuration model and those obtained from \algoname.
We focus on the configuration model as it is the standard reference model in network analysis~\cite{newman2018networks} and aligns with the focus of this paper on microcanonical ensembles.
Second, we analyze the extent to which the baseline algorithm \algofosdick differs from the color-aware algorithm \algomulti in their respective movements within the state space. 
Lastly, we show the scalability of both \algofosdick and \algomulti, particularly in relation to the number of vertex colors and the number of edges.

\mpara{Datasets.}
We consider $11$ real-world labeled networks, whose characteristics are summarized in \Cref{tbl:data} in \Cref{ax:data}.
% \textsc{Brexit}, \textsc{US-Elect}, 
% \textsc{Abortion}~\cite{garimella2017balancing}, \textsc{Twitter}~\cite{conover2011political}, \textsc{Obamacare}~\cite{garimella2017ebb},
% \textsc{Comb}, and \textsc{Guns}~\cite{garimella2018political}
% are retweet networks generated from tweets collected on various controversial topics. An edge exists between two users if one retweeted the other. Node colors indicate the side taken in the discussion, with a third label indicating neutrality.
% \textsc{Cite} and \textsc{Phy-Cit}
% % , and \textsc{Patents}
% ~\cite{preti2023maniacs} are citation networks: nodes represent publications, with node colors indicating Computer Science areas and the year of publication, respectively. 
% % In \textsc{Patents}, nodes are utility patents, with node colors indicating the year the patent was granted.
% \textsc{Trivago}~\cite{chodrow2021hypergraph} is network where nodes are accommodations, and edges connect accommodations visited by a user in the same browsing session. Node colors indicate the country where the accommodation is located.
% \textsc{Walmart}~\cite{Amburg-2020-categorical} is a co-purchase network where nodes are Walmart products, and edges connect products that were bought together. Node colors indicate the departments in which the products appear on \url{walmart.com}.

\mpara{Experimental Setup.}
All experiments are run on an Intel Xeon Silver 4210R CPU@2.40GHz running FreeBSD with $383$ GiB of RAM. 
We evaluate three sampling algorithms: the baseline color-agnostic algorithm \algofosdick, the color-aware algorithm \algomulti, and the traditional configuration model (\textsc{CM}), which samples from the state space of multigraphs with a prescribed degree sequence.
The code and the datasets used are available on GitHub.\footnote{\url{https://github.com/lady-bluecopper/Polaris}}

For the experiments aimed at the first goal, we allow $10$ Markov chains to evolve for $4000m$ iterations, where $m$ is the number of multiedges, recording the degree assortativity of the current state every $0.05m$ iterations.
%We also record the outcome of each iteration for the samplers and categorize them as: ($i$) the sampled DES is not a JDES, ($ii$) the sampled DES is a no-op JDES, ($iii$) the transition to the next state is accepted, or ($iv$) the transition is rejected.
For the experiments aimed at the other two goals, we generate $100$ independent samples by using each sampler for $m \log(m)$ iterations.

\subsection{Comparison with the Configuration Model}

\begin{figure}[tb]
      \centering
      \includegraphics[width=0.8\columnwidth]{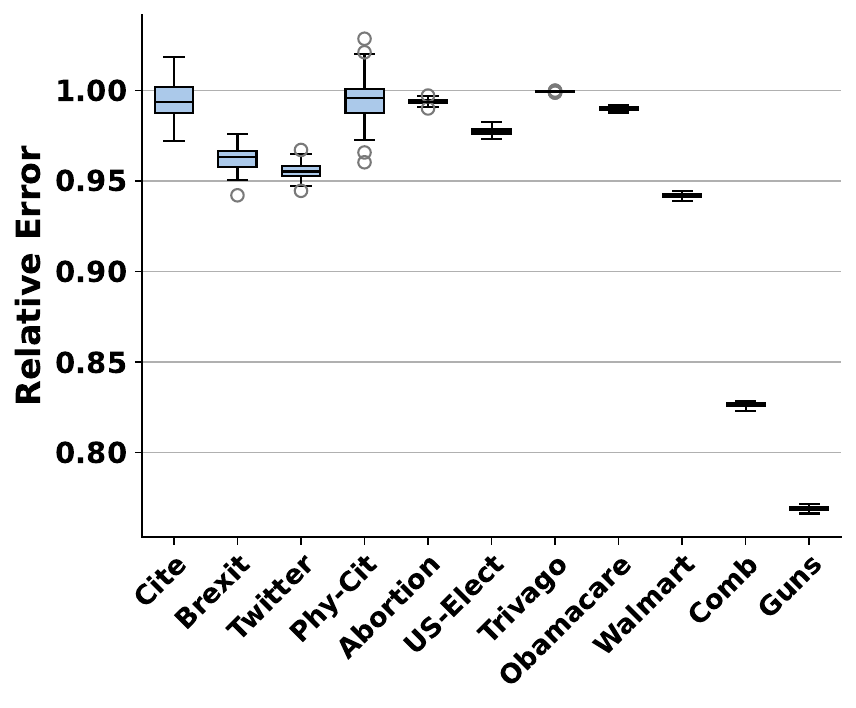}
      \vspace{-.5\baselineskip}
      \caption{Distribution of the relative errors of color assortativity for samples generated by \textsc{CM}, compared to the color assortativity of the observed datasets, for datasets of increasing size. Results are based on $100$ samples. Bars indicate one standard deviation.}
      \label{fig:ass_cm}
\end{figure}

\Cref{fig:ass_cm} shows that the color assortativity values of the multigraphs sampled by \textsc{CM} significantly diverge from those of the corresponding observed multigraphs, with relative errors close to $1$.
% This discrepancy is more pronounced in datasets with a larger number of colors. 
This discrepancy arises because \textsc{CM} disrupts the original correlations in the observed datasets, generating random graphs with low color assortativity. The effect is more pronounced in datasets with a larger number of colors or higher color assortativity, where the gap between the observed assortativity and that of the randomized graphs is larger. Consequently, we observe larger relative errors in datasets such as \textsc{Trivago} ($|\labelset|=160$), \textsc{Obamacare}, and \textsc{Abortion} (assortativity 0.95), and smaller errors in datasets such as \textsc{Comb} and \textsc{Guns} ($|\labelset|=2$ with assortativity values of 0.31 and 0.35, respectively).
This result proves that \textsc{CM} does not adequately capture the color assortativity present in the observed data.

\begin{figure}[t]
      \centering
      \includegraphics[width=0.8\columnwidth]{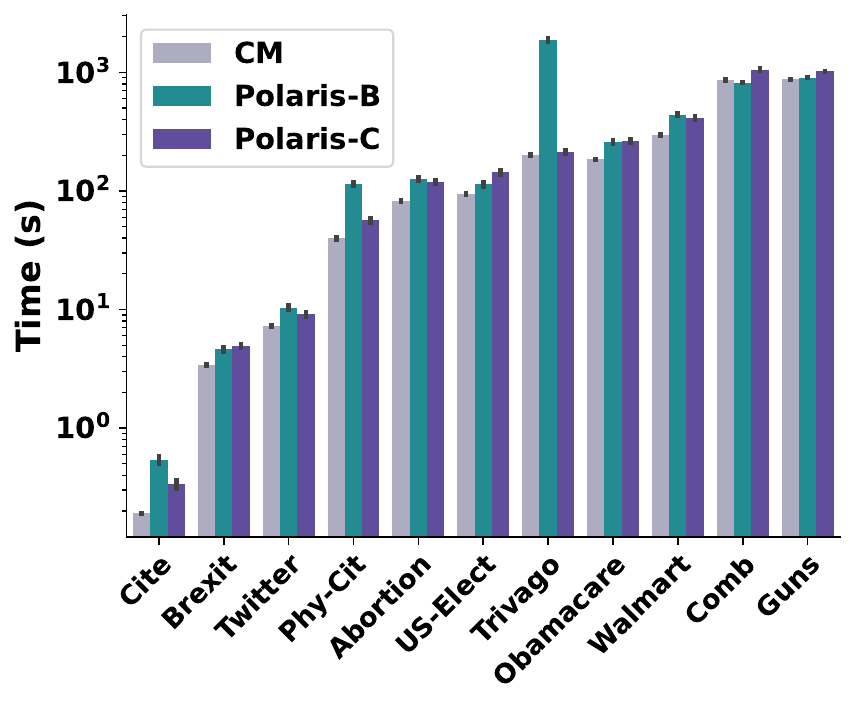}
      \vspace{-.5\baselineskip}
      \caption{Running time required by each sampler to perform $m\log(m)$ iterations, for datasets of increasing size. Results for $100$ samples. Bars indicate one standard deviation.}
      \label{fig:scalability}
\end{figure}

\Cref{fig:scalability} presents the running time of each sampler across the different datasets.
This plot highlights that, despite \algofosdick and \algomulti performing more complex operations and needing to update more quantities after each swap operation, their running time is similar to that of \textsc{CM}. 
However, the differences in performance become especially visible in datasets with a larger number of labels, such as \textsc{Trivago} and \textsc{Phy-Cit}. In these datasets, \algofosdick takes, on average, one order of magnitude longer to generate a sample compared to the other two samplers. 
As the number of colors increases, the likelihood that the sampled DES is not a JDES also increases, thus increasing the running time.
As a consequence, the algorithm must repeatedly sample new DESs until it finds one that is a JDES, which adds considerable overhead to the process.

\begin{figure}[t!]
      \centering
      \includegraphics[width=\columnwidth]{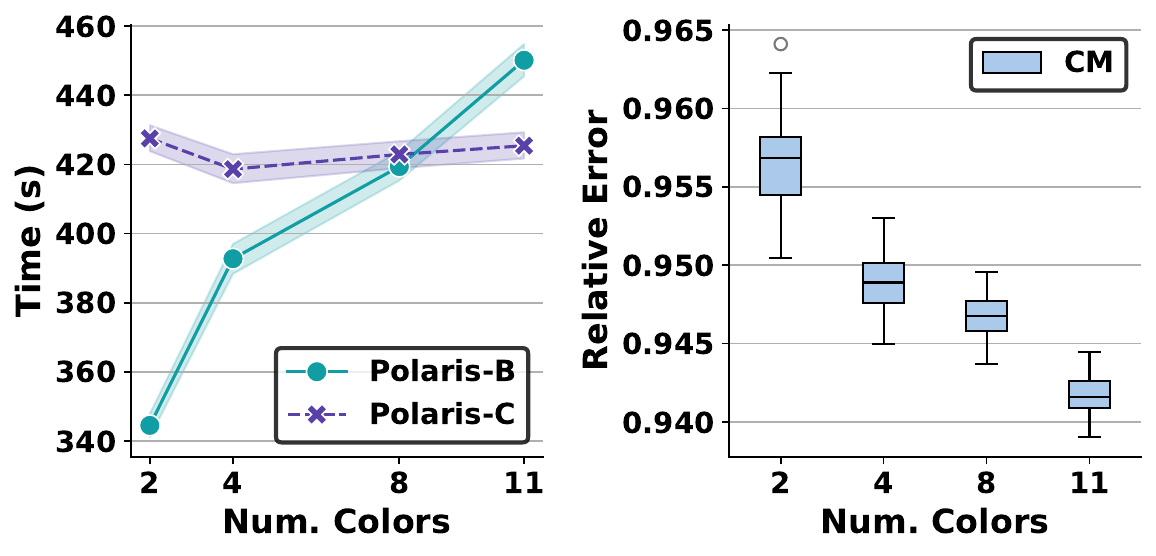}
      \vspace{-.5\baselineskip}
      \caption{Running time (left) required by \algofosdick and \algomulti to perform $m\log(m)$ iterations in different versions of \textsc{Walmart}, and distribution of the relative errors of color assortativity of the samples generated by \textsc{CM} (right). Results for $100$ samples.}
      \label{fig:labels}
\end{figure}

\Cref{fig:labels} presents the running time required by \algofosdick and \algomulti to perform $m \log(m)$ iterations on different versions of the \textsc{Walmart} dataset (left), and the distribution of the relative errors of color assortativity of the samples generated by \textsc{CM} (right).
Starting from the $11$ available colors (product categories), we cluster these colors to create new realistic sets of $2$, $4$, and $8$ colors.

The running time of \textsc{CM} is not affected by the number of colors, as it samples from a state space that is agnostic to vertex attributes.
Therefore we omit it from this plot.
Interestingly, the running time of \algomulti remains consistent across the different numbers of colors. This consistency is likely due to \algomulti maintaining a high acceptance rate, which produces a similar number of updates regardless of the number of colors.

In contrast, the running time of \algofosdick grows as the number of colors increases.
As already mentioned, a higher number of colors increases the probability that a sampled DES is not a JDES. 
Consequently, more DESs need to be drawn before finding one that is a JDES, which leads to an increased running time.

The figure also shows the distribution of the relative error of color assortativity values for the multigraphs generated by \textsc{CM}.
Again, we observe that the color assortativity of the sampled multigraphs significantly diverges from those of the original multigraphs.

\subsection{Performance Analysis of
\texorpdfstring{\algonamecollective}{\algonamecollectiveplain}}

\begin{figure}[t]
    \begin{subfigure}{\linewidth}
      \centering
      \includegraphics[width=\columnwidth]{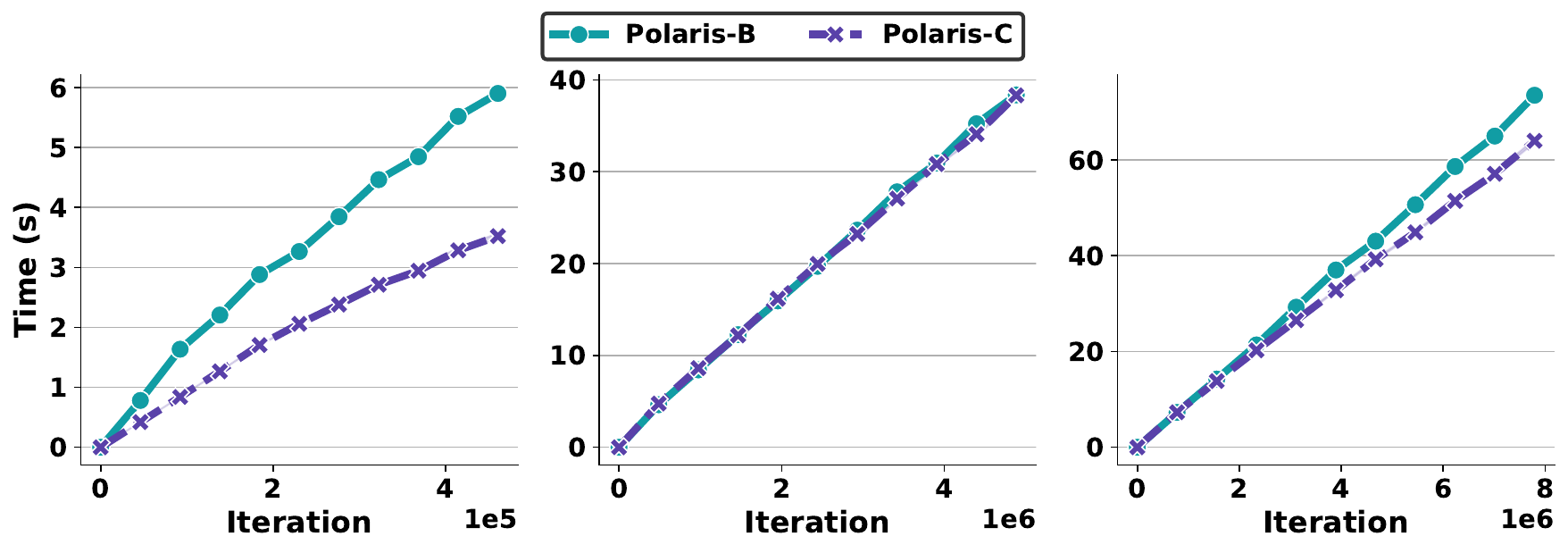}
    \end{subfigure}
    \begin{subfigure}{\linewidth}
      \centering
      \includegraphics[width=\columnwidth]{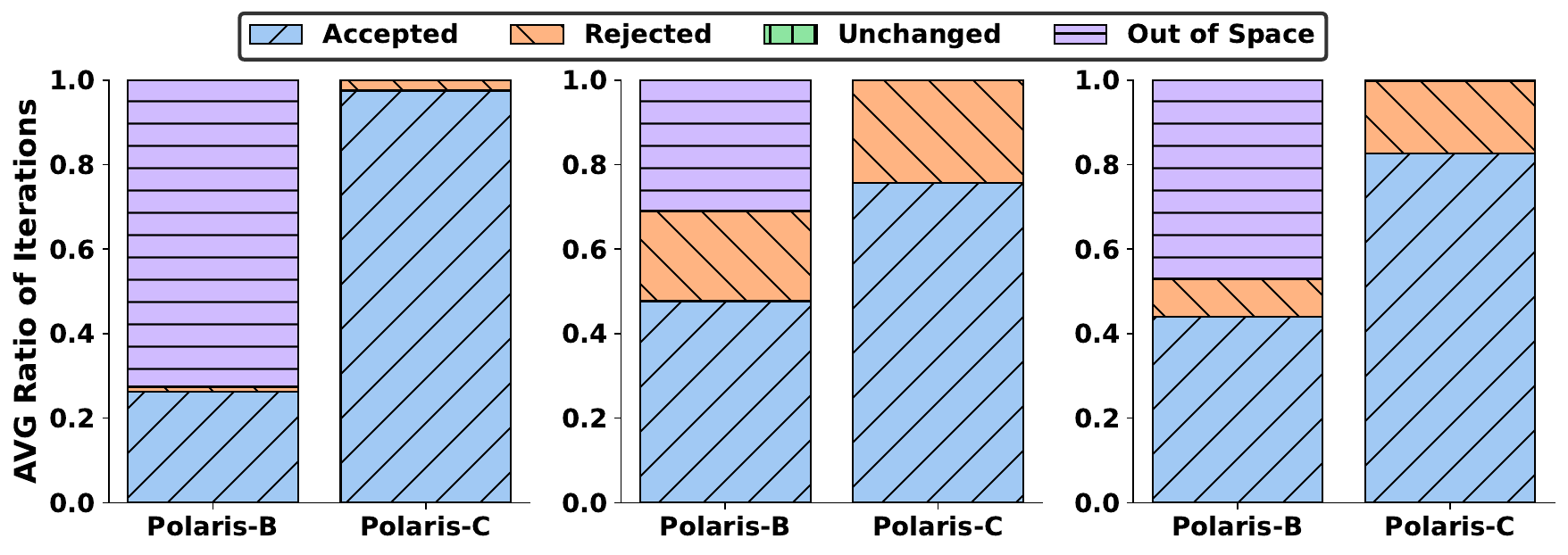}
    \end{subfigure}
    \begin{subfigure}{\linewidth}
      \centering
      \includegraphics[width=\columnwidth]{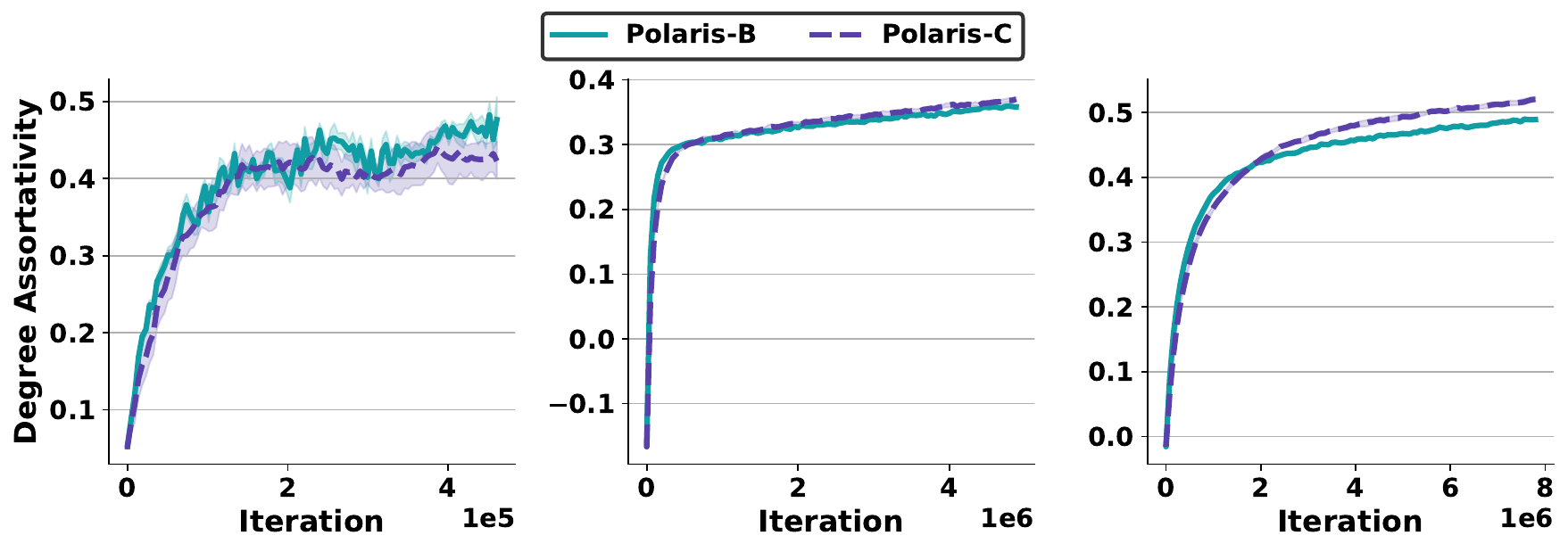}
    \end{subfigure}
    \vspace{-1.5\baselineskip}
    \caption{Running time (top), average ratio of iterations for each of the four possible outcomes (mid), and degree assortativity as a function of the number of iterations (bottom) for each sampler on \textsc{Cite} (left), \textsc{Brexit} (middle), and \textsc{Twitter} (right).}
    \label{fig:convergence}
\end{figure}

\Cref{fig:convergence} consists of three panels, each addressing a different aspect of the performance and behavior of \algofosdick and \algomulti, in three different datasets: \textsc{Cite}, \textsc{Brexit}, and \textsc{Twitter}.
The top panel shows the running time as a function of the number of iterations.
\algomulti is faster than \algofosdick in most cases.

The middle panel shows the fraction of iterations with four possible outcomes: ($i$) the sampled DES is not a JDES (\emph{Out of Space}), ($ii$) the DES is a no-op JDES (\emph{Unchanged}), ($iii$) an accepted transition to the next state (\emph{Accepted}), and ($iv$) a rejected transition (\emph{rejected}).
\algomulti avoids sampling DESs that are not JDES, thus having no \emph{Out of Space} outcomes. Additionally, \algomulti has a higher ratio of accepted transitions than \algofosdick, thus it explores the state space more extensively.
Due to frequent \emph{Out of Space} outcomes, \algofosdick has to frequently resample new DESs, leading to increased running times per iteration (as shown in the first panel). This effect becomes particularly evident as the number of colors increases.

The bottom panel illustrates the \emph{degree} assortativity of the states visited in the Markov chains produced by each sampler, which is often used as a measure of convergence for the chain~\citep{roy2019convergence}.
Both algorithms reach a plateau at nearly the same point, which suggests the states visited start to share similar characteristics. 
However, as shown in the first plot, \algomulti achieves this plateau faster.

% The third panel presents the results of the Gelman-Rubin diagnostic. To perform this diagnostic, we discarded the degree assortativity values collected in the first $100m$ iterations and then analyzed the values in the range $[t, 4000m]$ for each of the $10$ Markov chains, for each $t \geq 1000m$. The Gelmain-Rubin statistic compares the variance between chains to the variance within chains. A statistic value below 1.1 generally indicates convergence~\cite{gelman2004bayesian}. The results show that \algomulti reaches the desired value of 1.1 faster than \algofosdick, suggesting that \algomulti has a faster mixing time and achieves convergence more quickly. 

\begin{figure}[t]
    \begin{subfigure}{\linewidth}
      \centering
      \includegraphics[width=\columnwidth]{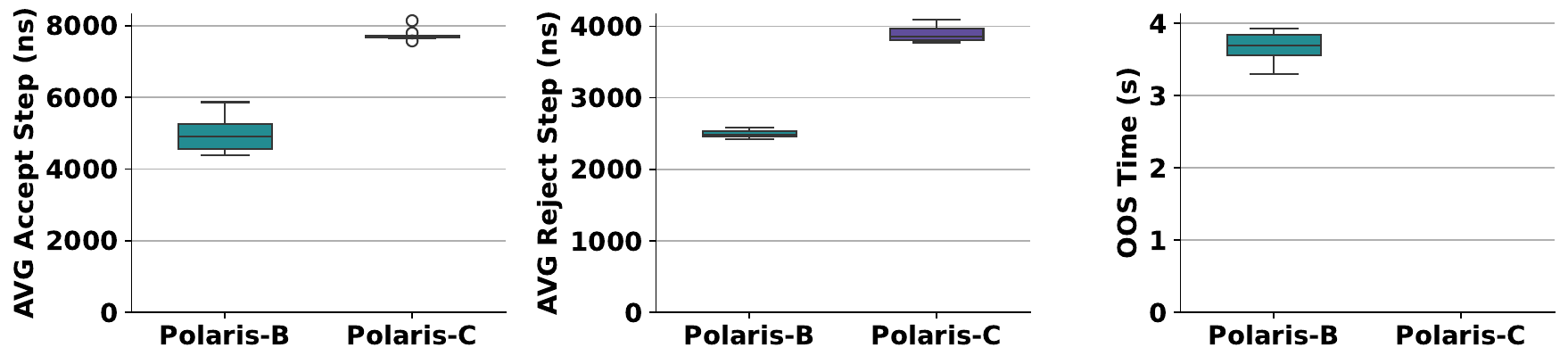}
      \caption{\textsc{Cite}}
    \end{subfigure}
    \begin{subfigure}{\linewidth}
      \centering
      \includegraphics[width=\columnwidth]{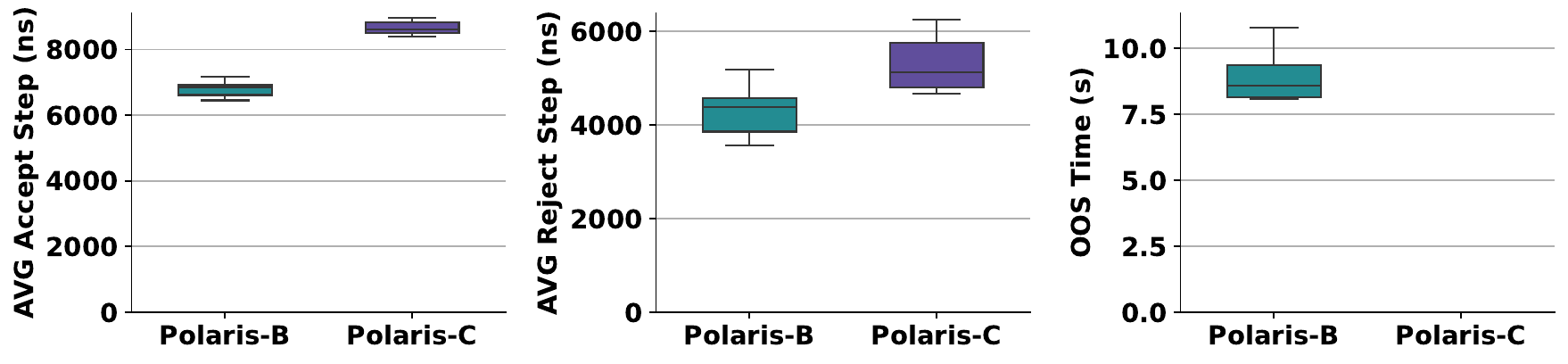}
      \caption{\textsc{Brexit}}
    \end{subfigure}
    \begin{subfigure}{\linewidth}
      \centering
      \includegraphics[width=\columnwidth]{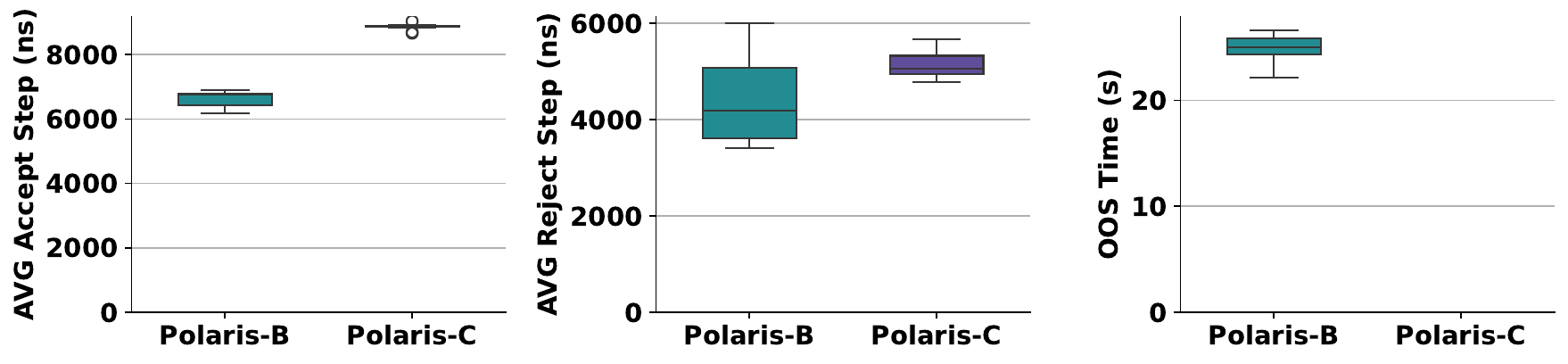}
      \caption{\textsc{Twitter}}
    \end{subfigure}
    \vspace{-1.5\baselineskip}
    \caption{Distribution of the average time required to perform a step where the transition to the next state is accepted (left) or rejected (middle). The right plots show the total time required to find a DES that is a JDES. Results for $10$ Markov chains.}
    \label{fig:avgsteps}
\end{figure}

\Cref{fig:avgsteps} provides a detailed analysis of the time required to perform a step in the sampling process, categorized by the type of outcome.
%The left and middle charts show the distribution of average step times for transitions based on their acceptance status.
Specifically, the left chart shows the times for transitions that are accepted, whereas the middle chart illustrates the times for transitions that are rejected. 
An accepted transition corresponds to the outcome \emph{Accepted}, while a rejected transition include the outcomes \emph{Rejected} and \emph{Unchanged}.
The right chart displays the distribution of total time for steps where the sampled DES is not a JDES (i.e., \emph{Out of Space}).
On average, accepting a transition is $2\times$ slower than rejecting it, because the algorithms must update the data structures that store the edges and their weights to maintain correctness after an accepted transition.

\algomulti has higher step times compared to \algofosdick as
\begin{squishlist}
\item \algomulti performs more computations even for rejected steps, as it must evaluate several quantities to compute the value of $\rho$. In contrast, \algofosdick performs fewer computations before rejection, and thus achieves lower running times;
\item when a transition is accepted, \algomulti needs to maintain additional data structures necessary to ensure that the sampled DES is always a JDES.
\end{squishlist}

Nonetheless, \algofosdick results in longer overall running times because it uses a considerable amount of time to find a DES that is a JDES, especially when the number of colors is higher.

% !TEX root =  ../main.tex
\section{Conclusion}\label{sec:concl}

We introduced \algoname, an ensemble of colored multigraphs with prescribed Joint Color Matrix. The JCM captures key properties relevant to the
study of polarized networks, such as color assortativity. 
We described two
efficient algorithms to sample from the space of such multigraphs according to
any user-specified probability distribution over this space. 
Our algorithms work
by running a Markov chain on the multigraph space, following the
Metropolis-Hastings approach for determining whether to accept the move to a
proposed neighbor of the current state. We conducted an extensive experimental
evaluation, showing the shortcomings of existing methods in capturing the color
assortativity, and assessing the performance of our algorithms across different
datasets in terms of scalability, runtime, and acceptance probability.

\rev{This work serves as an important first step toward analyzing polarization in real networks. While a comprehensive study of polarization lies beyond this paper's scope, the tools developed here lay the groundwork for future studies on this subject.}
% !TEX root =  ../main.tex
\section{Ethical considerations}\label{sec:ethical}

Our paper introduces a new method for assessing the statistical significance 
of the polarization structure discovered in online social networks.
The motivation of our work is the study of social phenomena, 
such as polarization, formation of echo chambers, opinion dynamics, and 
influence among individuals in online social networks.
As such, our work contributes to the growing field of computational social science, 
which in turn contributes to a better understanding of complex social behavior. 
Our emphasis in this work is on the design of new algorithms for efficient sampling 
from a novel network null model and the mathematical analysis of the algorithms and their properties. 
During our study, we did not perform any data collection, and our empirical evaluation 
uses benchmark graph datasets that are publicly available. 
For future studies and researchers who would like to apply our method
on new data collected from online social networks, 
we emphasize the importance of prioritizing ethical considerations 
to protect the privacy and rights of individuals. 
This involves obtaining informed consent where possible, 
anonymizing data to prevent the identification of users, 
applying the data minimization principle, and 
being mindful of the potential for harm in the analysis and dissemination of findings. 
It is also important to comply with platform policies and legal regulations, such as GDPR. 

\begin{acks}
  MR's work is supported by the \grantsponsor{NSF}{National Science
  Foundation}{https://www.nsf.gov} grants
  IIS-\grantnum[https://www.nsf.gov/awardsearch/showAward?AWD_ID=2006765]{NSF}{2006765},
  and CAREER-\grantnum[https://www.nsf.gov/awardsearch/showAward?AWD_ID=2238693]{NSF}{2238693}.
  AG's work is supported by the ERC Advanced Grant REBOUND (834862), 
  the EC H2020 RIA project SoBigData++ (871042), and
  the Wallenberg AI, Autonomous Systems and Software Program (WASP) 
  funded by the Knut and Alice Wallenberg Foundation.
\end{acks}

\bibliographystyle{ACM-Reference-Format}
\bibliography{biblio}

\appendix
% !TEX root =  ../main.tex
\section{Supplementary Material}
%
%\subsection{Reproducibility}\label{ax:code}

\subsection{Missing Proofs}\label{ax:proofs}

In this section we present the proofs that were not included in the main body of
the paper.

\begin{proof}[Proof of \cref{thm:connect}]
  For any two multigraphs $Q, R \in \nullset$, and any vertex $u \in V$, define
  \[
    \Delta_{Q, R}(u) \doteq \sum_{\ell \in
    \labelset}\abs{\nlneighs{u}{Q}{\ell} - \nlneighs{u}{R}{\ell}},
  \]
  i.e., as the sum of the absolute differences in the numbers of neighbors of
  $u$ with the same color across $Q$ and $R$. Also define
  \[
    \Delta(Q,R) \doteq \sum_{u \in V} \Delta_{Q, R}(u) \enspace.
  \]
  Let $G$ and $H$ be two distinct multigraphs in $\nullset$. We first build a
  sequence of JDESs to transform $G$ into a multigraph $\tilde{G} \in \nullset$
  such that $\Delta(\tilde{G}, H) = 0$, i.e., every vertex has in $\tilde{G}$
  exactly the same number of neighbors of each color as it has in $H$. Then, we
  construct a sequence of JDESs to transform $\tilde{G}$ into $H$.

  If $\Delta(G, H) = 0$, then let $\tilde{G} = G$. Otherwise, let $u$ be a
  vertex such that $\Delta_{G, H}(u) > 0$. From this fact and since
  $\degree{u}{G} = \degree{u}{H}$ (as the two multigraphs have the same degree
  sequence), then, there are distinct $j,k \in \labelset$ s.t.\ %
  $\nlneighs{u}{G}{j} < \nlneighs{u}{H}{j}$ and $\nlneighs{u}{G}{k} >
  \nlneighs{u}{H}{k}$.

  It follows from the first inequality and the fact that $G$ and $H$ have
  the same JCM, that there must be a vertex $v \neq u$ with $\labelf{v} =
  \labelf{u}$ and such that $\nlneighs{v}{G}{j} > \nlneighs{v}{H}{j}$.

  Let $w \in \lneighs{u}{G}{k}$ and $z \in \lneighs{v}{G}{j}$. These two
  vertices necessarily both exist because $\nlneighs{u}{G}{k} >
  \nlneighs{u}{H}{k} \ge 0$, and $\nlneighs{v}{G}{j} > \nlneighs{v}{H}{j} \ge
  0$.

  Since $\labelf{u} = \labelf{v}$, we can construct the JDES $\swap{(u,w),
  (v,z)}{(u,z), (v,w)}$ that transforms $G$ into some $T \in \nullset$. This
  JDES falls in case 4C of the characterization of JDESs. It holds:
  \begin{itemize}
    \item $\Delta_{T, H}(u) = \Delta_{G, H}(u) - 2$;
    \item $\Delta_{T, H}(v)$ is either equal to $\Delta_{G, H}(v)$
      or to $\Delta_{G, H}(v) - 2$, as $\abs{\nlneighs{v}{T}{\ell} -
      \nlneighs{v}{H}{j}} = \abs{\nlneighs{v}{G}{j} - \nlneighs{v}{H}{j}} -
      1$, and $\abs{\nlneighs{v}{T}{k} -
      \nlneighs{v}{H}{k}}$ is $\abs{\nlneighs{v}{G}{k} - \nlneighs{v}{H}{k}} \pm
      1$;
    \item $\Delta_{T, H}(q) = \Delta_{G, H}(q)$ for every other vertex $q
      \in V \setminus \{u,v\}$, as $w$ and $z$ exchanged neighbors with the
      same color $\labelf{u}$, and every other vertex has no change in its
      neighborhood.
  \end{itemize}
  Therefore, it holds $\Delta(T, H) \le \Delta(G, H) - 2$. By repeatedly
  applying the procedure above, we eventually obtain a graph $\tilde{G} \in
  \nullset$ such that $\Delta(\tilde{G}, H) = 0$.

  Now, if $\tilde{G} = H$, we are done. Otherwise, for any multigraph $Q = (V,
  E, \labelset, \labelfsym) \in \nullset$, and any unordered pair $(\ell, r)$ of
  colors from $\labelset$ (potentially $\ell = r$), define $V_{\ell,r} = \{ v
  \in V \suchthat \labelf{v} \in \{\ell , r\}\}$, and define the subgraph
  $Q_{\ell,r}$ of $Q$ as the multigraph
  \[
    Q_{\ell,r} = (V_{\ell,r}, \mset{(u,v) \in E \suchthat \labelf{u} = \ell
    \wedge \labelf{v} = r}, \{\ell, r\}, \labelfsym_{|V_{\ell,r}}),
  \]
  i.e., the subgraph of $Q$ that contains only the vertices with color $\ell$ or
  $r$, and only the edges that have one endpoint with color $\ell$ and one
  endpoint with color $r$. Clearly, if $\ell=r$, $Q_{\ell,\ell}$ is the subgraph
  of $Q$ induced by the vertices with color $\ell$, but if $\ell \neq r$,
  $Q_{\ell,r}$ is not the subgraph of $Q$ induced by the vertices with color
  $\ell$ or $r$, as in this case $Q_{\ell,r}$ does not include any eventual
  self-loop over its vertices, as its edges are all and only the bichrome edges with
  one vertex of color $\ell$ and the other vertex of color $r$. The edge sets
  of the various $Q_{\ell,r}$ form a partitioning of the edge set $E$ of $Q$.

  For $\ell \in \labelset$, consider $\tilde{G}_{\ell,\ell}$ and
  $H_{\ell,\ell}$. These multigraphs have the same set of vertices, and the same
  degree sequence, as $\Delta(\tilde{G}, H) = 0$. Since their vertices all have
  color $\ell$,  any DES from $\tilde{G}_\ell$ is a JDES, falling in either case
  1, 2A, 2B, 3A, 3B, or 4C. A classic result of graph theory (see, e.g.,
  \citep[Lemma 2.14]{fosdick2018configuring}) states that there is a sequence of
  double edge swaps connecting any multigraph with vertices all with the same
  color to any other multigraph with the same degree sequence as the starting
  multigraph. Thus there is a sequence of JDESs that connects
  $\tilde{G}_{\ell,\ell}$ to $H_{\ell,\ell}$. For any arbitrary ordering
  $\ell_1,\ell_2,\dotsc,\ell_{\card{\labelset}}$ of the colors in $\labelset$,
  we can then consider the sequence of JDESs obtained by concatenating the
  sequences of JDESs connecting $\tilde{G}_{\ell_i,\ell_i}$ to
  $H_{\ell_i,\ell_i}$, and apply the JDESs in the resulting sequence, starting
  from $\tilde{G}$, to obtain $\check{G}$ with $\check{G}_{\ell,\ell} =
  H_{\ell,\ell}$ for $\ell \in \labelset$.

  The same approach can also be used for $\check{G}_\mathrm{\ell,r}$
  ($=\tilde{G}_{\ell,r}$) and $H_{\ell,r}$ for $\ell \neq r$. They have the same
  set of vertices and the same degree sequences, and any DES is a JDES, falling
  in either case 2C, 3D, or 4A. Every bipartite multigraph can be transformed,
  through a sequence of DES, into any other bipartite multigraph with the same
  set of vertices and the same degree sequence. Thus there is a sequence of
  JDESs that transforms $\check{G}_\mathrm{\ell,r}$ ($=\tilde{G}_{\ell,r}$) into
  $H_{\ell,r}$, for $\ell \neq r$. Using the same arbitrary ordering
  $\ell_1,\ell_2,\dotsc,\ell_{\card{\labelset}}$ of the colors in $\labelset$,
  we can then consider the ordering $(\ell_1,\ell_2), (\ell_1, \ell_3),
  \allowbreak \dotsc, (\ell_1, \ell_{\card{\labelset}}),\allowbreak
  (\ell_2, \ell_3), \dotsc, (\ell_{\card{\labelset} -1},
  \ell_{\card{\labelset}})$ of the unordered pairs of different colors, and
  consider the sequence of JDESs obtained by concatenating the sequences of
  JDESs connecting $\check{G}_{\ell_i,\ell_j}$,
  $i < j$, to $H_{\ell_i,\ell_j}$, to obtain $H$.

  We have thus built a sequence of JDESs that starts at $G$, goes through
  $\tilde{G}$ and $\check{G}$, and reaches $H$. Since every JDES is reversible,
  our proof is complete.
\end{proof}

\begin{proof}[Proof of \cref{thm:aperiod}]
  Assume that only the first condition holds.

  If the edges fall in Cases 1, 2B, 2C, 2D, 3B, 3C, 3D, or 3E, then at least one
  of the JDESs involving them is a no-op, so the state graph has a self-loop on
  state $G$, and is therefore aperiodic.

  If the edges fall in Cases 2A or 3A, then applying either of the equivalent
  JDESs involving the edges leads to a state $H \neq G$ where the
  new edges fall in case 2B or either 3B or 3C respectively, meaning that the
  state graph has a self-loop on $H$, thus it is aperiodic.

  If the edges fall in case 4C,\footnote{The proof for this case is the same as
  the proof for the aperiodicity of the state graph by DESs \citep[Lemma
  3]{fosdick2017configuring}.} the state graph has a cycle of length two,
  because every JDES is reversible. It also has a cycle of length three, by
  applying the following sequence of JDESs: $\swap{(u,w), (v,z)}{(u,v),(w,z)}$,
  $\swap{(u,v),(w,z)}{(u,z),(v,w)}$, and $\swap{(u,z),(v,w)}{(u,w),(v,z)}$. The
  greatest common divisor of the lengths of these two cycles is one, thus the
  state graph is aperiodic.

  Assume now that only the second condition holds.\footnote{The proof for this
    case is the same as the proof for the aperiodicity of the state graph by DES
    when considering only bipartite graphs as the states \citep[Sect.\
  4.2.1]{ponocny2001nonparametric}.} The state graph has a cycle of length two,
  because every JDES is reversible. It also has a cycle of length three, by
  applying the following sequence of JDESs: $\swap{(u,z),(w,x)}{(u,x),(w,z)}$,
  $\swap{(u,v),(w,z)}{(u,z),(w,v)}$, $\swap{(u,x),(w,v)}{(u,v),(w,x)}$. The
  greatest common divisor of the lengths of these two cycles is one, thus the
  state graph is aperiodic.
\end{proof}

\begin{proof}[Proof of \cref{thm:algofosdickcorrect}]
  As shown in \cref{thm:connect,thm:aperiod}, the state graph is
  strongly-connected and aperiodic. For every $G,H \in \nullset$ s.t. $H$ is a
  neighbor of $G$, \algofosdick\ the probability $\neighprob{G}{H}$ of proposing
  $H$ when the state of the chain is $G$ is the same as in \citep[Algorithm
  3]{fosdick2017configuring}, and thus so is the ratio $\rho \doteq
  \neighprob{H}{G} / \neighprob{G}{H}$ used by \algofosdick, and therefore the
  acceptance probability $\acceptprob{G}{H} \doteq \min\{1, \rho
  \nullprob(H)/\nullprob(G)\}$. Thus, \algofosdick\ follows the MH approach, and
  the Markov chain it runs has stationary distribution $\nullprob$.
\end{proof}

We now give the proof to \cref{thm:algomulticorrect}. In the proof, we use the
following immediate facts.

\begin{fact}\label{fact:nocase0}
  For any $\ell \in \labelset$ and any pair of distinct edges $(u,w),$ $(v,z)$ $ \in
  E_{G,\ell}$, there is always a JDES from $G$ involving these two edges, as
  $\{\labelf{u},\labelf{w}\} \cap \{\labelf{v},\labelf{z}\} \neq \varnothing$,
  i.e., Case 0 in the characterization of JDESs never holds.
\end{fact}

\begin{fact}\label{fact:oneortwosets}
  Let $(u,w), (v,z) \in E_{G,\ell}$. If $\card{\{\labelf{u}, \labelf{w},
  \labelf{v}, \labelf{z}\}} \neq 2$, there exists no $r \in \labelset$, $r \neq
  \ell$, such that $(u,w), (v,z) \in E_{G,r}$. Otherwise, there is exactly one
  such $r$, as $\{\labelf{u}, \labelf{w}, \labelf{v}, \labelf{z}\} = \{\ell,
  r\}$.
\end{fact}

\begin{fact}\label{fact:backandforth}
  If $G, H \in \nullset$ are neighbors and the one or two JDESs that transform
  $G$ into $H$ fall in case \textsf{XY} from the classification above, then the
  one or two JDESs that transform $H$ into $G$ fall in the same case
  \textsf{XY}, with the following exceptions:
  \begin{description}
    \item[\textsf{XY}=2A:] the only JDES from $H$ to $G$ falls in case 2B;%
    \item[\textsf{XY}=2B:] the two equivalent JDESs from $H$ to $G$ fall in case
      2A;%
    \item[\textsf{XY}=3A:] the only JDES from $H$ to $G$ falls in either case 3B
      or 3C, depending on whether the three involved vertices have all the same
      color (3B) or not (3C);
    \item[\textsf{XY}=3B:]  the only JDES from $H$ to $G$ falls in case 3A;%
    \item[\textsf{XY}=3C:] the only JDES from $H$ to $G$ falls in case 3A.%
  \end{description}
\end{fact}

\begin{fact}\label{fact:constantsize}
  Let $G \neq H \in \nullset$. For any $\ell \in \labelset$, it holds
  $\card{E_{G,\ell}} = \card{E_{H,\ell}}$, i.e., the size of these sets is
  constant across all multigraphs in $\nullset$.
\end{fact}

The above fact does not imply $E_{G,\ell} = E_{H,\ell}$.

\begin{proof}[Proof of \cref{thm:algomulticorrect}]
  Let $G \in \nullset$ be the current state of the Markov Chain, i.e., the
  value taken by the variable $G$ at the beginning of some iteration of the loop
  (\cref{algline:loopstart}). We aim to show that \algomulti\ follows the
  Metropolis-Hastings (MH) approach. To this end, we need to show that the
  acceptance probability $\acceptprob{G}{H}$ is computed according to the MH
  approach, for any neighbor $H$ of $G$.

  Let $\ell$ take value $r \in \labelset$ (\cref{algline:labsample}), and let
  $(u,w)=(a,b)$ and $(v,z)=(c,d)$ be the edges sampled from $E_{G,r}$
  (lines~\ref{algline:edge1sample}--\ref{algline:edge2sample}). It holds from
  \cref{fact:nocase0} that here is always a JDES from $G$ involving these edges
  (i.e., Case 0 from the characterization never holds). Let
  $\swap{(a,b),(c,d)}{(a,c),(b,d)}$  be a JDES involving these edges, and let
  $H \in \nullset$ be the multigraph resulting from applying the JDES to $H$. We
  now consider the different cases for the JDES.%

  In Cases 1, 2C, 2D, 3D, or 3E:\@ it holds $H = G$, and indeed the
  algorithm does not update $G$
  (lines~\ref{algline:case1},\ \ref{algline:continue2C}, and~\ref{algline:continue3D}), i.e., the state of the
  Markov chain is unchanged, as required by MH.%

  For all other cases, it will be $H \neq G$. To assess $\acceptprob{G}{H}$, we
  need to study the neighbor proposal probabilities $\neighprob{G}{H}$ and
  $\neighprob{H}{G}$. More specifically, we study the value of the variable
  $\rho$ set by \algomulti, and we show that is always set to $\neighprob{H}{G}
  / \neighprob{G}{H}$. The correctness of the algorithm then follows from this
  fact and the fact that \algomulti\ decides whether to accept $H$ by comparing
  a real drawn uniformly at random from $[0,1]$ to the value $\rho \nullprob{H}
  / \nullprob{G}$, i.e., it accepts $H$ with probability $\acceptprob{G}{H} =
  \min \{1, \rho \nullprob{H} / \nullprob{G} \}$, as required by MH.%

  Consider the random variables $\ell$, $(u,w)$, $(v,z)$, and $\mathsf{jdes}$
  used by the algorithm, and defined the following events:
  \begin{description}
    \item[$\mathsf{E}_\text{l}$:] $\ell=r$ such that $(a,b),(c,d) \in E_{G,r}$;
    \item[$\mathsf{E}_\text{e}$:] $((u,w) = (a,b) \wedge (v,z) = (c,d)) \vee
      ((u,w) = (c,d) \wedge (v,z) = (a,b))$;
    \item[$\mathsf{E}_\text{j}$:] $\mathsf{jdes} =
      \swap{(a,b),(c,d)}{(a,c),(b,d)}$.
  \end{description}
  It clearly holds $\neighprob{G}{H} = \Pr(\mathsf{E}_\text{j})$. Using the
  law of total probability, we can write:
  \[
    \neighprob{G}{H} = \Pr(\mathsf{E}_\text{j}) = \Pr(\mathsf{E}_\text{l})
    \Pr(\mathsf{E}_\text{e} \mid \mathsf{E}_\text{l}) \Pr(\mathsf{E}_\text{j}
    \mid \mathsf{E}_\text{e}) \enspace.
  \]
  We now analyze $\mathsf{jdes}$, using the characterization
  of the JDESs.

  \spara{Case 3A, 3B, 3C, 4A:} for these cases, it holds $\{\labelf{a},
      \labelf{b}\}$ $\cap$ $\{\labelf{c}, \labelf{d}\}$ $=$ $\{r\}$ for some $r \in
      \labelset$. Thus, \cref{fact:oneortwosets} tells us that $(a,b)$ and
      $(c,d)$ appear together only in $E_{G,r}$, hence $\Pr(\mathsf{E}_\text{l})
      = 1/\card{\labelset}$. Also, the two edges are not copies of the same
      multiedge, hence
      \begin{equation}\label{eq:pree3A}
        \Pr(\mathsf{E}_\text{e} \mid \mathsf{E}_\text{l}) = \frac{\mul{(a,b)}{G}
        \mul{(c,d)}{G}}{\card{E_{G,\ell}} (\card{E_{G,\ell}} - 1)} \enspace.
      \end{equation}
      Finally, it clearly holds $\Pr(\mathsf{E}_\text{j} \mid
      \mathsf{E}_\text{e}) = 1$. Thus,
      \begin{equation}\label{eq:neighprob3A}
        \neighprob{G}{H} = \frac{\mul{(a,b)}{G} \mul{(c,d)}{G}}{\card{\labelset}
        \card{E_{G,\ell}} (\card{E_{G,\ell}} - 1)} \enspace.
      \end{equation}
      We know from \cref{fact:backandforth} that the JDES
      $\swap{(a,c),(b,d)}{(a,b),(c,d)}$ from $H$ to $G$ would fall in a case
      among those we are considering here, so we can obtain $\neighprob{H}{G}$
      from \cref{eq:neighprob3A} as
      \[
        \neighprob{H}{G} = \frac{\mul{(a,c)}{H} \mul{(b,d)}{H}}{\card{\labelset}
        \card{E_{H,\ell}} (\card{E_{H,\ell}} - 1)} \enspace.
      \]
    We can then use \cref{fact:constantsize} and the fact that $\mul{(a,c)}{H} =
    \mul{(a,c)}{G} + 1$ and $\mul{(b,d)}{H} = \mul{(b,d)}{G} + 1$ to rewrite the
    above as
    \begin{equation}\label{eq:neighprob3AHG}
      \neighprob{H}{G} = \frac{(\mul{(a,c)}{G} + 1) (\mul{(b,d)}{G} + 1)}{\card{\labelset}
      \card{E_{G,\ell}} (\card{E_{G,\ell}} - 1)} \enspace.
    \end{equation}
    It follows that \algomulti, on
    \cref{algline:rho3A,algline:rho3B,algline:rho3C,algline:rho4A} sets $\rho$
    to the ratio $\neighprob{H}{G} / \neighprob{G}{H}$, as requested.

  \spara{Case 4C:} in this case, at least one of the edges is monochrome, so
      $\Pr(\mathsf{E}_\text{l}) = 1/\card{\labelset}$.  The edges are not
      copies, so $\Pr(\mathsf{E}_\text{e})$ is as in \cref{eq:pree3A}. But it holds
      $\Pr(\mathsf{E}_\text{j} \mid \mathsf{E}_\text{e}) = 1/2$, since
      $\mathsf{jdes}$ is equally likely to be either of the two JDESs involving
      the sampled edges, only one of which transforms $G$ into $H$. Thus,
      $\neighprob{G}{H}$ equals the r.h.s.\ of \cref{eq:neighprob3A} multiplied
      by $1/2$. The JDES from $H$ to $G$ falls in this same case, so we can
      proceed as in the previous case, and $\neighprob{H}{G}$ equals the r.h.s.\
      of \cref{eq:neighprob3AHG} multiplied by $1/2$. Thus, \algomulti\ sets
      $\rho$ on \cref{algline:rho4C1,algline:rho4C2} to the ratio
      $\neighprob{H}{G} / \neighprob{G}{H}$, as required.

  \spara{Case 2A:} this case is partly similar to Cases 3A, 3B, 3C, 4A, and the
      value for $\neighprob{G}{H}$ is the same as in \cref{eq:neighprob3A},
      noting that, in this case, the JDES from $G$ to $H$ can be written as
      $\swap{(a,a),(c,c)}{(a,c),(a,c)}$. On the other hand, we know from
      \cref{fact:backandforth} that the JDES $\swap{(a,c),(a,c)}{(a,a),(c,c)}$
      from $H$ to $G$ falls in case 2B. This case is analyzed next, and the
      proposal probability $\neighprob{H}{G}$ can be obtained from
      \cref{eq:neighprob2B}. By using \cref{fact:constantsize}
      and the fact that $\mul{(a,c)}{H} = \mul{(a,c)}{G} + 2$, similarly to how
      we proceeded in the above cases, we obtain
      \[
        \neighprob{H}{G} = \frac{(\mul{(a,c)}{G} + 2) (\mul{(a,c)}{G} +
        1)}{\card{\labelset} \card{E_{G,\ell}} (\card{E_{G,\ell}} - 1)}
        \enspace.
      \]
      Thus, \algomulti\ sets $\rho$ on \cref{algline:rho2A} to the ratio
      $\neighprob{H}{G} / \neighprob{G}{H}$,
      as required.

  \spara{Case 2B:} this case is also partly similar to the first we analyzed,
      except that the two edges are copies of the same multiedge $(a,b)$, hence
      \[
        \Pr(\mathsf{E}_\text{e} \mid \mathsf{E}_\text{l}) = \frac{\mul{(a,b)}{G}
        (\mul{(a,b)}{G} - 1)}{\card{E_{G,\ell}} (\card{E_{G,\ell}} - 1)}
      \]
      and
      \begin{equation}\label{eq:neighprob2B}
        \neighprob{G}{H} = \frac{\mul{(a,b)}{G} (\mul{(a,b)}{G} - 1)}{\card{\labelset}
        \card{E_{G,\ell}} (\card{E_{G,\ell}} - 1)} \enspace.
      \end{equation}
      From \cref{fact:backandforth}, we know that the JDES from $H$
      to $G$ would fall in case 2A, thus the proposal probability
      $\neighprob{H}{G}$ can be obtained from \cref{eq:neighprob3A}, following a
      process similar to the one we described in case 2A, and resulting in
      \[
        \neighprob{H}{G} = \frac{(\mul{(a,a)}{G} + 1) (\mul{(b,b)}{G} + 1)}{\card{\labelset}
      \card{E_{G,\ell}} (\card{E_{G,\ell}} - 1)} \enspace.
      \]
      Therefore, \algomulti\ sets $\rho$ on \cref{algline:rho2B} to
      the ratio $\neighprob{H}{G} / \neighprob{G}{H}$, as required.

  \spara{Case 4B:} in this case, the two edges $(a,b)$ and $(c,d)$ are both
      bichrome with the same two colors $r'$ and $r''$, thus
      \cref{fact:oneortwosets} tells us that they appear together in both
      $E_{G,r'}$ and $E_{G,r''}$, hence $\Pr(\mathsf{E}_\text{l}) =
      2/\card{\labelset}$. The edges are not copies, so
      \[
        \Pr(\mathsf{E}_\text{e} \mid \mathsf{E}_\text{l}) = \frac{\mul{(a,b)}{G}
          \mul{(c,d)}{G}}{\card{E_{G,r'}} (\card{E_{G,r'}} - 1)} + \frac{\mul{(a,b)}{G}
        \mul{(c,d)}{G}}{\card{E_{G,r''}} (\card{E_{G,r''}} - 1)} \enspace.
      \]
      It holds
      $\Pr(\mathsf{E}_\text{j} \mid \mathsf{E}_\text{e}) = 1$, thus
      \[
        \neighprob{G}{H} = \frac{2}{\card{\labelset}} \left( \frac{\mul{(a,b)}{G}
          \mul{(c,d)}{G}}{\card{E_{G,r'}} (\card{E_{G,r'}} - 1)} + \frac{\mul{(a,b)}{G}
        \mul{(c,d)}{G}}{\card{E_{G,r''}} (\card{E_{G,r''}} - 1)} \right) \enspace.
      \]
      The JDES from $H$ to $G$ also falls in case 4B, per
      \cref{fact:backandforth}. Using \cref{fact:constantsize}, and the
      multiplicities of the edges in $G$ to express those in $H$, we obtain
      \begin{align*}
        \neighprob{H}{G} = \frac{2}{\card{\labelset}} & \left(
        \frac{(\mul{(a,d)}{G} + 1)(\mul{(c,b)}{G}+1)}{\card{E_{G,\ell}}
        (\card{E_{G,\ell}} - 1)} \right. \\
        & \left. + \frac{(\mul{(a,d)}{G} +
      1)(\mul{(c,b)}{G}+1)}{\card{E_{G,\ell'}} (\card{E_{G,\ell'}} - 1)} \right)
        \enspace.
      \end{align*}
      Once again, \algomulti\ clearly sets $\rho$ to the ratio $\neighprob{H}{G}
      / \neighprob{G}{H}$ on \cref{algline:rho4B}.
\end{proof}

\subsection{Datasets}\label{ax:data}
We consider $11$ real-world labeled networks, whose characteristics are summarized in \Cref{tbl:data}.
\textsc{Brexit}, \textsc{US-Elect}, 
\textsc{Abortion}~\cite{garimella2017balancing}, \textsc{Twitter}~\cite{conover2011political}, \textsc{Obamacare}~\cite{garimella2017ebb},
\textsc{Comb}, and \textsc{Guns}~\cite{garimella2018political}
are retweet networks generated from tweets collected on various controversial topics. An edge exists between two users if one retweeted the other. Node colors indicate the side taken in the discussion, with a third label indicating neutrality.
\textsc{Cite} and \textsc{Phy-Cit}
% , and \textsc{Patents}
~\cite{preti2023maniacs} are citation networks: nodes represent publications, with node colors indicating Computer Science areas and the year of publication, respectively. 
% In \textsc{Patents}, nodes are utility patents, with node colors indicating the year the patent was granted.
\textsc{Trivago}~\cite{chodrow2021hypergraph} is network where nodes are accommodations, and edges connect accommodations visited by a user in the same browsing session. Node colors indicate the country where the accommodation is located.
\textsc{Walmart}~\cite{Amburg-2020-categorical} is a co-purchase network where nodes are Walmart products, and edges connect products that were bought together. Node colors indicate the departments in which the products appear on \url{walmart.com}.

\begin{table}[h!]
  \small
  \caption{Dataset characteristics: number of vertices, number of edges, average and median degree, and average and median color frequency.}
  \label{tbl:data}
  \resizebox{\columnwidth}{!}{%
  \begin{tabular}{lrrrrrrr}
    \toprule
     Dataset & $|V|$ & $|E|$ & $|\labelset|$ & $\widehat{\degree{u}{}}$ & $\widetilde{\degree{u}{}}$ & $\widehat{|V^\ell|}$ & $\widetilde{|V^\ell|}$\\
    \midrule
    \textsc{Cite} & 3264 & 4611 & 6 & 2.83 & 2.00 & 0.17 & 0.18 \\
    \textsc{Brexit} & 22745 & 48830 & 2 & 4.29 & 1.00 & 0.50 & 0.50 \\
    \textsc{Twitter} & 22405 & 77920 & 3 & 6.96 & 1.00 & 0.33 & 0.32 \\
    \textsc{Phy-Cit} & 30501 & 347268 & 11 & 22.77 & 14.00 & 0.09 & 0.11 \\
    \textsc{Abortion} & 279505 & 671144 & 2 & 4.80 & 1.00 & 0.50 & 0.50 \\
    \textsc{US-Elect} & 23832 & 845152 & 3 & 70.93 & 3.00 & 0.33 & 0.25 \\
    \textsc{Trivago} & 172738 & 1327092 & 160 & 15.37 & 6.00 & 0.01 & 0.00 \\
    \textsc{Obamacare} & 334617 & 1511670 & 2 & 9.04 & 1.00 & 0.50 & 0.50 \\
    \textsc{Walmart} & 88860 & 2267396 & 11 & 51.03 & 18.00 & 0.09 & 0.05 \\
    \textsc{Comb} & 677753 & 6666018 & 2 & 19.67 & 1.00 & 0.50 & 0.50 \\
    \textsc{Guns} & 632659 & 7478993 & 2 & 23.64 & 1.00 & 0.50 & 0.50 \\
    % \textsc{Patents} & 2745762 & 13965410 & 4 & 10.17 & 7.00 & 0.25 & 0.24 \\
    \bottomrule
  \end{tabular}
  }
\end{table}

\subsection{Computational Complexity}
\algomulti has an initialization step for the data structures used to ensure we only sample pairs of edges that form a JDES. This initialization phase has a time complexity of $\mathcal{O}(n + m)$ and requires $\mathcal{O}(n + m)$ space, where $n$ is the number of vertices and $m$ is the number of edges. The space is used to store information such as node degrees, node colors, the edge list, and the edge subsets $E_{G,\ell}$ for each color $\ell$. Since each edge can belong to up to two of these subsets, they occupy at most $2m$ space in total.
Each step of the algorithm then takes constant time, and we perform $s$ steps. In our experiments, we follow previous works and set $s = m \log(m)$ as ~\cite{viger2005efficient}.
The number of colors $|\labelset|$ does not affect the time complexity, as \algomulti directly samples pairs of edges that can form a JDES. 
In contrast, \algofosdick is affected by $|\labelset|$. As $|\labelset|$ increases, so does the number of possible combinations of node colors, thus reducing the probability that two sampled edges belong to the same subset $E_{G,\ell}$. Each time this condition is not met, \algofosdick must resample two new edges, which increases its running time.

\end{document}